\documentclass[%
 reprint,
 amsmath,amssymb,
 aps,
 floatfix,
]{revtex4-1}

\usepackage{graphicx}
\graphicspath{{images/}}
\usepackage{dcolumn}
\usepackage{bm}

\begin{document}

\preprint{APS/123-QED}

\title{Synchronization of Conservative Parallel Discrete Event Simulations on a Small-World Network}

\author{Liliia Ziganurova $^{1,2}$}
 \altaffiliation {ziganurova@gmail.com}
\author{Lev N. Shchur $^{1,2}$}%
 \email{lev.shchur@gmail.com}
\affiliation{%
 $^1$ Science Center in Chernogolovka, 142432 Chernogolovka, Russia \\
 $^2$ National Research University Higher School of Economics,  101000 Moscow, Russia
 }%

\date{\today}

\begin{abstract}
We examine the question of the influence of sparse long-range communications
on the synchronization in parallel discrete event simulations (PDES). We
build a model of the evolution of local virtual times (LVT) in a
conservative algorithm including several choices of local links.
All network realizations belong to the small-world network class. We find
that synchronization depends on the average shortest path of the network.
The time profile dynamics are similar to the surface profile growth, which
helps to analyze synchronization effects using a statistical physics
approach. Without long-range links of the nodes, the model belongs to the
universality class of the Kardar--Parisi--Zhang equation for surface growth.
We find that the critical exponents depend logarithmically on the fraction
of long-range links. We present the results of simulations and discuss our
observations.
\end{abstract}

\maketitle


\section{\label{sec:level1}Introduction}

Progress in computation hardware in the last decade has been mainly in the
direction of multicore/distributed systems. It is a big challenge to use
modern hardware effectively, and creating a single program able to
orchestrate a huge number of nodes and cores is not
trivial~\cite{joubert2016parallel,zhu2015high}.

Here, we discuss the problem of synchronization within one family of
parallel simulations. The class of considered systems comprises very many
individual elements interacting asynchronously with each other, and events
occur at some discrete instants. Simulating such systems using sequential
algorithms requires a vast amount of processing time and memory. The method
for simulating systems on parallel/distributed computers, which allows
implementing a faithful synchronization, is called parallel discrete event
simulation (PDES)~\cite{fujimoto1990parallel,fujimoto2016research}.

The simulation technique is used in many areas of physics; examples include
simulation of granular dynamics~\cite{richardson2011numerical}, kinetic
Monte Carlo simulations~\cite{nilmeier2014rigorous}, and simulation of 3D
sintering~\cite{garcia2010parallel}. It is proved to work on millions of
cores~\cite{oppelstrup2016spock}.

It was shown in~\cite{korniss2000massively} that evolution of the simulated
time profile in PDES is analogous to the evolution of nonequilibrium surface
growth. A model of the time profile evolution was proposed, and in the case
where the processing elements (PEs) communicate only with neighbors, such a
model can be mapped on the (1+1)-dimensional Kardar--Parisi--Zhang (KPZ)
equation~\cite{kardar1986dynamic}. This finding helps to understand the
synchronization problem in the language of statistical physics. For example,
(i) the positivity of the profile speed is mapped on the property of
deadlock absence, and (ii) the evolution of the profile width, which is
described with the KPZ critical exponents, reflects the desynchronization
of the PEs. Taking these into account, we mainly use the language of
statistical physics instead of computational science in what follows.

Assuming local communication between the PEs restricts the model application
to a relatively small number of applications. Generally, long-range
communications between processor elements do occur in simulations. It is
reasonable to investigate a more realistic link topology.

Here, we consider PDES on small-world (SW) networks~\cite{watts1998collective}.
The current state of research with SW networks is presented in
Section~\ref{sec:SW}. The main topological feature of a SW network is that
for relatively small amount of long-range links, the average distance
changes from a linear to a logarithmic dependence on the system
size~\cite{newman2010networks}. Clearly, this should drastically change the
behavior of the whole system. It was found in~\cite{korniss2003suppressing,guclu2006synchronization}
that random long-range links between PEs strongly influence synchronization
properties and the scalability of PDES. A synchronization scheme with
additional long-range links introduces a relaxation term in the evolution
of the virtual time profile. This term implies the absence of
large-amplitude long-wavelength modes~\cite{guclu2006synchronization} in the
surface. Consequently, the average width of the profile becomes finite,
while the average progress rate remains a nonzero constant in the limit of
infinite system size. In other words, (i) introducing long-range links does
not change the important property of the local conservative algorithm, the
deadlock absence; (ii) the long-range links increases the synchronization of
simulations. It was also found that the average width in sufficiently large
systems is proportional to the correlation length $\xi(p)$, and $\xi(p)\sim
p^{-0.84}$, where $p$ is the probability of the long-range interactions.

We construct the topology of the communications between PEs in the framework
of the SW approach~\cite{watts1998collective}. The concentration $p$ of
long-range communications is the main parameter in our research. We find
that the clustering coefficient value does not qualitatively influence the
development of surface growth. The quantitative change of the rate of
surface growth and the surface width behavior is independent of the local
connectivity. For this, we analyze networks with only nearest neighbors,
with nearest and next-nearest neighbors, and so on. We thus find some
universal properties. Our main conclusion is that the average length of the
network and number of local connections govern the surface growth dynamics.
The average length is a function of the parameter $p$ and is known to behave
logarithmically for values that are not too small. It is important that we
do not change the update scheme of the conservative PDES algorithms as in
the papers~\cite{korniss2003suppressing, guclu2006synchronization}. Our
purpose is to investigate how the SW topology of the communication links
influences the synchronization properties of PDES, i.e., the statistical
properties of the surface growth~\cite{foo1}.

We build our model on two types of SW networks. Both have a small average
shortest path (the main criterion of ``small-worldness''). One SW
realization has a zero clustering coefficient (the second feature of SW
networks~\cite{watts1999small} is a nonzero clustering coefficient). We find
that the average speed profile decreases slowly as the parameter $p$
increases and the speed is always positive. The average profile width
becomes finite in the limit of an infinite system size in accordance with
the result in~\cite{guclu2006synchronization}. Another new result here is an
estimate of the dependence of the growth exponent $\beta$ on the SW
parameter $p$: the dependence is logarithmic.

This paper is organized as follows. In section \ref{sec:PDES}, we describe
the conservative model for PDES \cite{korniss2000massively}.
Section~\ref{sec:SW} contains a detailed description and an analysis of SW
topologies. In section \ref{sec:SWmodel}, we describe our one-dimensional
SW scheme and present the results. In section \ref{sec:Local}, we analyze
the dependence of the measured quantities on the number of local links. In
section \ref{sec:Results}, we summarize our work and discuss the results.

\section{Basic Conservative PDES scheme}
\label{sec:PDES}

Parallel discrete event simulation is a subclass of parallel simulation
where changes in the components of the system from one state to another
occur instantaneously. These changes are called \textit{events}. The
system being simulated is divided into disjoint subsystems. Subsystems
are processed by PEs, which are hardware dependent and may be a computing
node, CPU, core, or thread. In the simplest case, each PE carries only one
site of the underlying system (e.g., one spin in a magnetic model). The
important feature of PDES is that the PEs communicate with each other
asynchronously and via messages. Each PE progresses at its own pace and
has its own simulated time, also called the local virtual time
(LVT)~\cite{jefferson1985virtual}. Different synchronization schemes are
possible for preserving the causality of computations~\cite{shchur2004evolution}.
We focus our discussion on a conservative algorithm, which avoids the
possibility of any type of causality error by checking every causality
relation at each update attempt \cite{fujimoto1990parallel}.

The \textit{model} of the time profile evolution regards LVT as Poisson
arrivals. In the basic one-dimensional case, the network topology is a
ring~\cite{korniss2000massively}, and PEs hence interact only with nearest
neighbors. Let $N$ be the number of PEs and $t$ be the number of parallel
steps. The set of LVTs $\{\tau_i(t)\}_{i=1}^{N}$ constitutes the
\textit{virtual time profile}. At each time step, only those PEs whose
LVTs are not larger than the LVTs of their nearest neighbors may increment
their LVTs by an exponentially distributed random value. These PEs are
said to be \textit{active}. Otherwise, if the LVT of a PE is larger than
the LVT of some neighbor, this PE is not updated and is said to be
\textit{passive}. The relative amount of active PEs (those simulating
system evolution) is called the \textit{utilization}
\begin{equation}
 \langle u(t,N)\rangle= \biggl \langle \frac{N(t)_{\mathrm{active}}}{N}\biggr \rangle
\label{eq:u}
\end{equation}
and is an important characteristic of the evolution of the LVT profile. The average $\langle\,\cdot\,\rangle$ is taken
over many independent realizations. In
the basic conservative scheme, the utilization at the given instant is equal
to the density of local minima of the profile, $N_{\min}/N$. The value of
the utilization can be used as a measure of algorithm effectiveness.

The second important observable is the spread or \textit{width} of the LVT
profile, defined as
\begin{equation}
\langle w^2(N,t)\rangle=\biggl\langle\frac{1}{N}\sum_{i=1}^{N}[\tau_i(t)-\overline\tau(t)]^2\biggr\rangle,
\label{eq:w2}
\end{equation}
where $\overline\tau(t)=\frac{1}{N}\sum_{i=1}^{N}\tau_i(t)$ is the mean
height of the time profile.

As the number of PEs in a parallel architecture increases to hundreds of
thousands, a fundamental question of the scalability of the underlying
algorithm emerges. To be scalable, a PDES algorithm must have the following
properties: (1) the LVT profile should progress on average with a nonzero
rate, and (2) the width of the profile should be bounded by a constant as
the number of PEs goes to infinity. A PDES algorithm is said to be
\textit{fully scalable} if both conditions are satisfied \cite{toroczkai2003virtual}.
It is interesting that the scalability of computations is defined in the
limit of an infinite system size. This is one more analogy with the
corresponding physical system for which the thermodynamic limit is reached
in the same limit.

We briefly recall the main results of a study of the basic conservative
scheme~\cite{korniss2000massively}. The LVT profile width increases with
time and then saturates to the steady-state regime after some time
$t_\times$. Before saturation, the width grows as $\langle w^2(t)\rangle\sim
t^{2\beta}$, where $\beta=0.326(5)$. In the steady state, the width is
stationary and depends on the system size $\langle w^2_{\infty}\rangle\sim
N^{2\alpha}$, $\alpha=0.49(1)$. The two values of the exponents $\alpha$
and $\beta$ are close to those of the KPZ universality class \cite{kardar1986dynamic},
$\alpha=1/2$ and $\beta=1/3$. The estimate of the utilization of the
algorithm (measure of the algorithm effectiveness) given in \cite{shchur2004evolution}
is $\langle u\rangle_\infty=0.246410(7)$. Therefore, the
basic conservative algorithm is \textit{computationally scalable} in one
dimension because the average utilization is greater than zero. But the
width of the LVT profile diverges as the number of nodes increases, which
means that the PEs became less synchronized. Therefore, the conservative
PDES algorithm is not fully scalable. In other words, the algorithm is still
applicable for any large system (it somehow progresses in time with positive
utilization) although it becomes less and less effective as the number of
PEs increases because the PEs become more and more desynchronized as the
simulation progresses (the width of the time increases with the number of
PEs).

\section{Small-world networks}
\label{sec:SW}

Small-world networks comprise a class of networks usually characterized by
a small average shortest path length and a high degree of clustering. These
properties are observed in many real technological, biological, social, and
information networks. There is no rigid definition of ``small-worldness,''
and different criteria for classifying networks into regular, SW, and random
classes have been proposed during the last
decade~\cite{watts1999small, humphries2008network, barranca2015low, borassi2015hyperbolicity}.

\begin{figure}[ht!]
\center \begin{minipage}{0.45\linewidth}
\center{\includegraphics[width=1\linewidth]{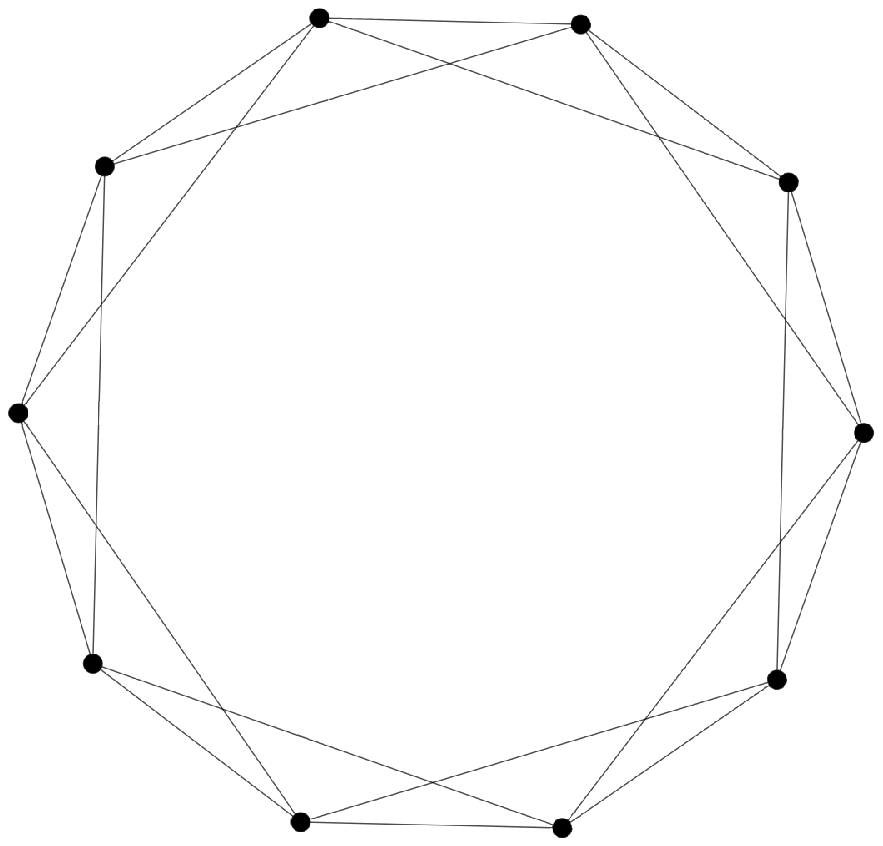}} a) \\
\end{minipage}
\hfill
\begin{minipage}{0.45 \linewidth}
\center{\includegraphics[width=1\linewidth]{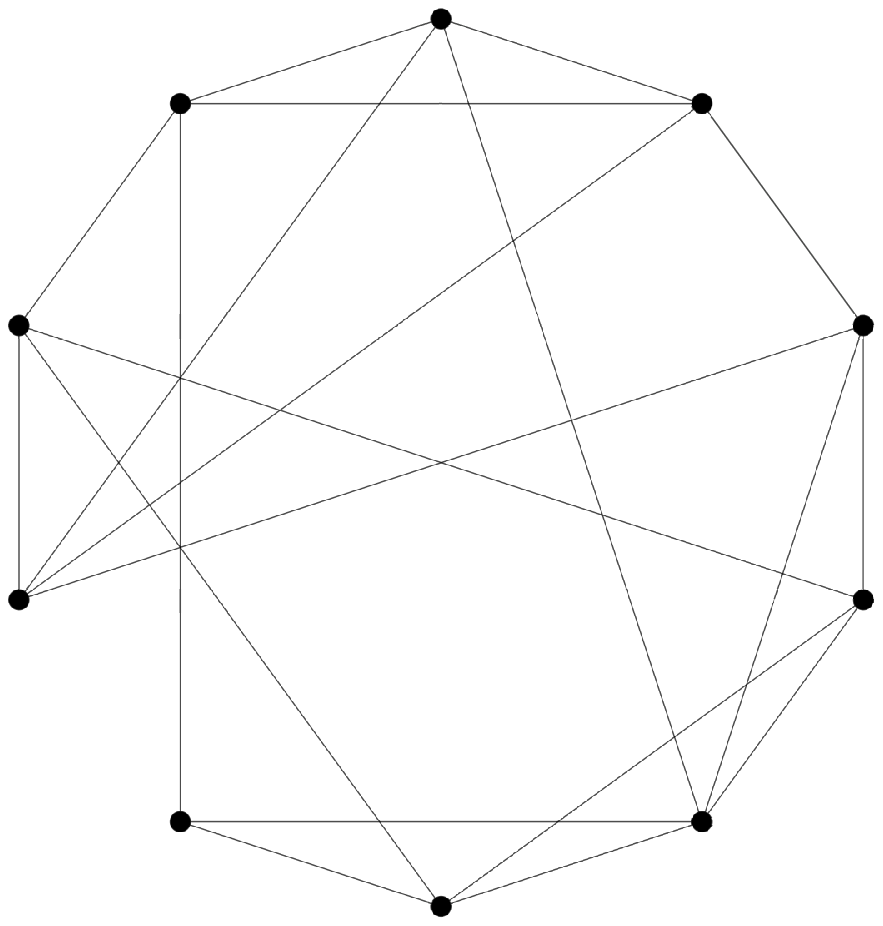}} b) \\
\end{minipage}
\caption{(a) A one-dimensional lattice with each site connected to its $2k$
neighbors with periodic boundary conditions: in this case $k=2$, (b) The
Watts and Strogatz model, where a small fraction of the links are rewired to
new sites chosen randomly, or {\em R--k2} model.}
\label{pic:k2}
\end{figure}

For precision, we first give some basic definitions and notations.
We consider a one-dimensional lattice with periodic boundary
conditions, where each node is connected with $2k$ neighbors
(Fig.~\ref{pic:k2}a). We call two nodes {\em neighbors} if there is an
edge between them. The total number of nodes is denoted by $N$. We also need
a parameter $p$, which can be interpreted as a degree of randomness. We
consider two structural properties of networks, the average shortest path
and the clustering coefficient.

\begin{figure}[ht!]
\center \begin{minipage}{0.45\linewidth}
\center{\includegraphics[width=1\linewidth]{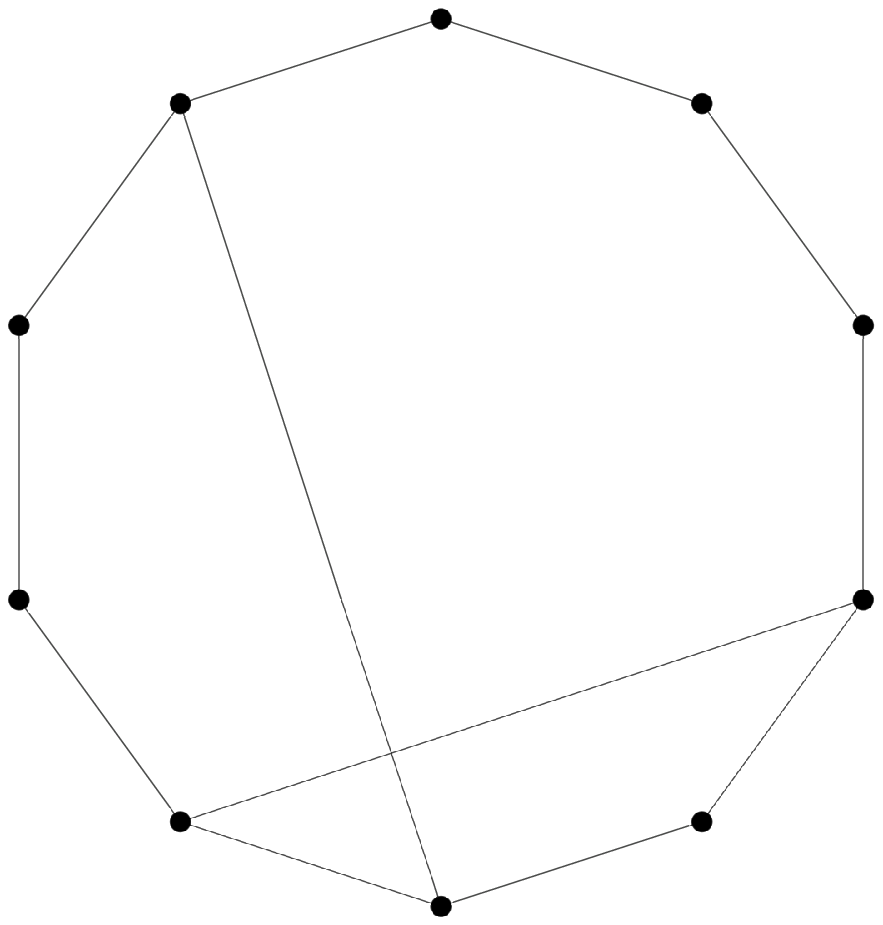}} a) \\
\end{minipage}	
\hfill
\begin{minipage}{0.45 \linewidth}
\center{\includegraphics[width=1\linewidth]{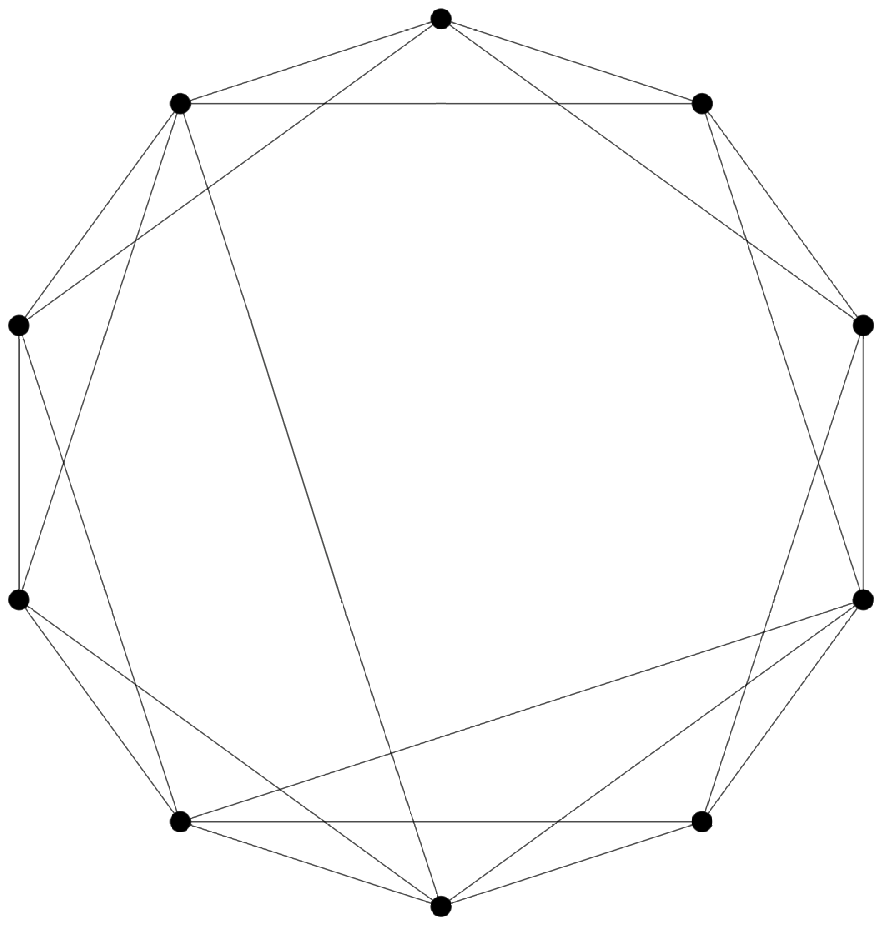}} b) \\
\end{minipage}
\caption{SW networks with each site connected to its $2k$ neighbors and a
small fraction of links added above the regular lattice with periodic
boundary conditions: (a) {\em A--k1}, and (b) {\em A--k2}.}
\label{pic:k1}
\end{figure}

There are several ways to construct networks with long-range links. Given a
one-dimensional lattice with each node connected to $2k$ closest nodes (see
Fig.~\ref{pic:k2}a), each edge of the graph is randomly rewired with
probability $p$, i.e., one end of the edge is moved to a node chosen at
random from the rest of the lattice nodes (see Fig.~\ref{pic:k2}b). Another
way to build a network is by adding links with probability $p$ above the
regular lattice (see Fig.~\ref{pic:k1}).

We conduct our study on three different networks based on both constructions
described above. For simplicity, we give a short code names to the networks:
{\em ``A (add) or R (rewrite) -- parameter $k$''}. The construction
algorithms are:
\begin{enumerate}
\item {\em A--k1}. (1) Start with a ring lattice with $N$ nodes where
each node is connected to its $k{=}1$ closest nodes. (2) Randomly add
exactly $pN$ edges above the regular lattice (Fig.~\ref{pic:k1}a).
\item {\em A--k2}. (1) Start with a ring lattice with $N$ nodes where
each node is connected to its $k{=}2$ closest nodes (Fig.~\ref{pic:k2}a).
(2) Randomly add exactly $pN$ edges above the regular lattice
(Fig.~\ref{pic:k1}b).
\item {\em R--k2}. (1) Start with a ring lattice with $N$ nodes where
each node is connected to its $k{=}2$ closest nodes. (2) Randomly choose
exactly $pN$ edges and rewrite them randomly (Fig.~\ref{pic:k2}b).
\end{enumerate}

The parameter $p$ thus can be regarded as the average number of random
long-range links per node.


\paragraph {Average shortest path.} The average shortest path $l(N,p)$ is
defined as
\begin{equation}
l(N,p)=\frac{1}{N(N-1)}\sum_{i\not=j}d_{ij},
\end{equation}
where $d_{ij}$ is a {\em chemical distance}~\cite{barrat2000properties}, the
minimum number of nodes between vertices $i$ and $j$.

In regular lattices, the average shortest path grows linearly with the
system size:
$$
l(N,0)=\frac{N(N+2k-2)}{4k(N-1)} \sim N/4k.
$$
For $p=1$ the length $l(N,1)$ grows as:
$$
l(N,1) \sim \frac{\ln(N)}{\ln(2k-1)}.
$$
For SW networks, we have the scaling
relation~\cite{newman2000mean, newman1999renormalization}
\begin{equation}
l(N,k,\tilde p)=\frac{N}{k}f((\tilde pk)^{1/d}N),
\label{lengthscaling}
\end{equation}
where $d$ is a lattice dimension, $\tilde p = p/k$ is the concentration of long-range links normalized with the number of local connections,  and $f(x)$ is a universal scaling function,
\begin{equation}
f(x)=\begin{cases}\mathrm{const}&\text{if }x\ll1,\\
\ln(x)/x&\text{if }x\gg1.\end{cases}
\end{equation}

The above relation indicates a crossover transition between regular and
SW networks. The number of rewired or added links ($pN$) must be small but
finite. The regime with $x=pN\ll1$ is not easily attained in practice for
networks of a finite size $N$.

\paragraph{Clustering coefficient.} The clustering coefficient $C(p)$
quantifies a ``cliquishness'' of a network. It is defined as follows. Let
$c_i$ be the number of neighbors of a node $i$. Node $i$ can have at most
$c_i(c_i-{1})/2$ possible links between all its neighbors. Let $N_i$ be the
actual number of such links. Then the local clustering is
$C_i=N_i/(c_i(c_i-1)/2)$, and the clustering coefficient $C(p)$ is the
average local clustering over all $N$ nodes~\cite{barrat2000properties}.

The clustering coefficient of a regular lattice is high:
$C(0)=3(k-1)/2(2k-1)$. In contrast, random networks are not clustered:
$C_{\mathrm{rand}}\sim k/N$. There are several analytic estimates of the
clustering coefficient of SW
networks~\cite{barrat2000properties, newman2002structure, watts1999small}.

For example, Barrat and Weight~\cite{barrat2000properties} derived
expression~(\ref{eq:Barrat}) for the clustering coefficient based on the
reasoning that the local clustering coefficient in the SW network remains
the same as in a regular lattice if all three edges connecting the node to
its two neighbors and the neighbors between themselves are not rewired.
This happens with probability $(1-\tilde p)^3$:
\begin{equation}
C(\tilde p)\approx C(0)(1-\tilde p)^3.
\label{eq:Barrat}
\end{equation}

Watts in his book~\cite{watts1999small} used a more complex analysis of
clustering phenomena and derived his expression~(\ref{eq:Watts}) via the
effective local degree and the effective global degree (for more detail,
see Chapter 4 in the book~\cite{watts1999small}):
\begin{equation}
C(\tilde p)\approx\frac{\frac{3}{4}(1-\tilde p)^2(2k-\frac{2}{3})-(1-\tilde p)}{2k-1}.
\label{eq:Watts}
\end{equation}

\begin{figure}[ht!]
\center{\includegraphics[width=1\linewidth]{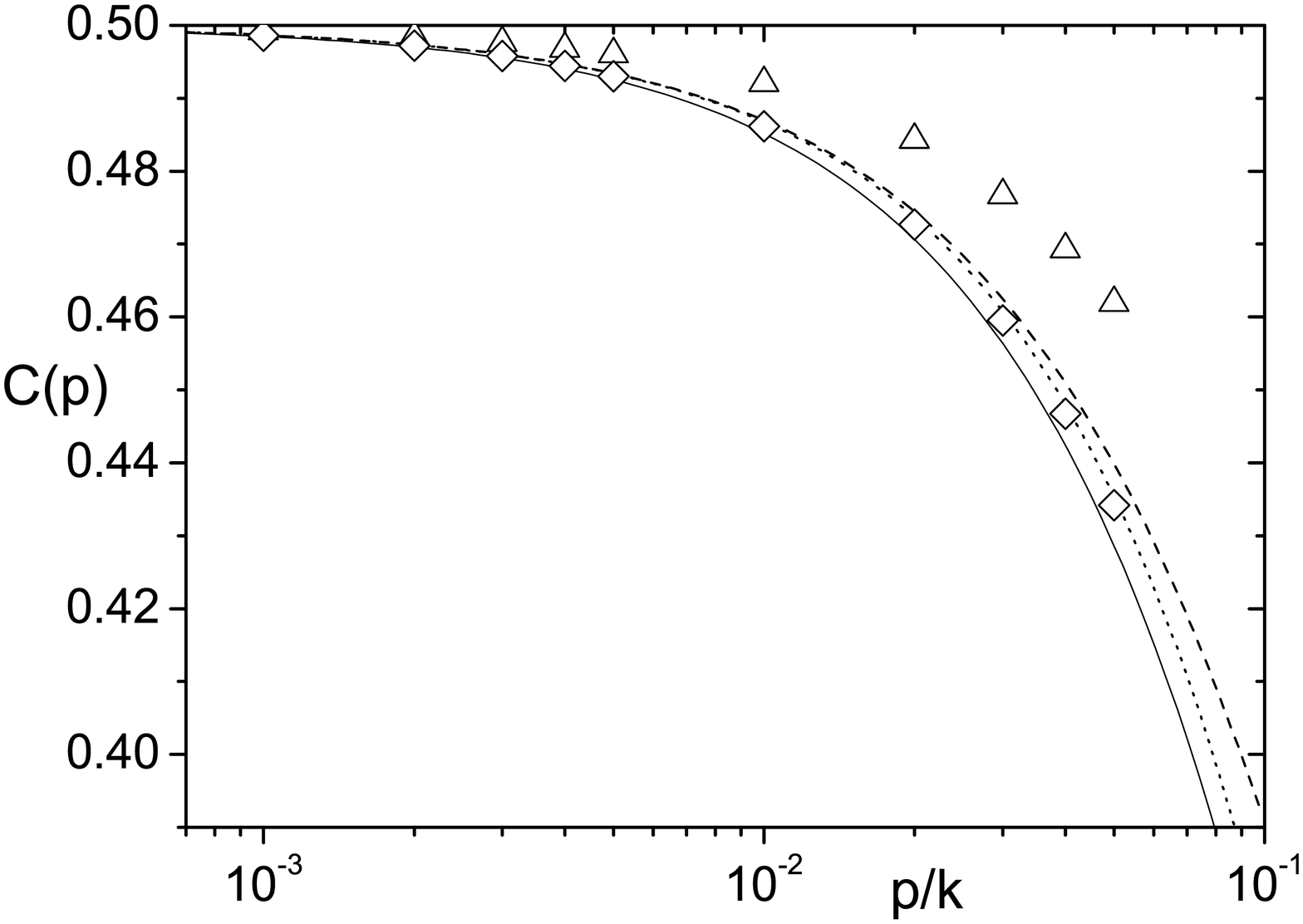}}
\caption{The clustering coefficient $C(p)$ of SW networks as a function of the
parameter $\tilde p = p/k$: triangles are {\em A--k2}, dimonds are
{\em R--k2}, the solid line indicates equation~(\ref{eq:Barrat}), the dotted line indicates equation~(\ref{eq:Watts}), and the dashed
line indicates equation~(\ref{eq:Newman}). Error bars are of the symbol size.}
\label{pic:clustering}
\end{figure}

One more formula is Newman's~\cite{newman2002structure} equation
\begin{equation}
C( \tilde p)\approx\frac{3k(k-1)}{2k(2k-1)+8\tilde pk^2+4\tilde p^2k^2}.
\label{eq:Newman}
\end{equation}
In this formula, $C(p)$ decreases slowly with $p$ and hence remains
sufficiently high for a small amount of long-range links
(Fig.~\ref{pic:clustering}).
These three formulas are derived for the SW-networks constructed by rewiring links,
it is {\em R--k2} in our case. It is seen from Fig.~\ref{pic:clustering}, that clustering
coefficient for the network {\em A--k2} follow the formulas only for a very small value of $p$.
For a SW network {\em A--k1} with $p=0$, we have $C(0)=0$. Adding long-range links
increases the probability of a nonzero clustering coefficient in such a network, namely,
the clustering coefficient $C(p)\sim p/N^2$ for {\em A--k1}.

The informal SW definition at the beginning of this section can now be
formulated more precisely: ``a SW graph is a large-$N$, sparsely connected,
decentralized graph ($N\gg k\gg1$) with a characteristic path length close
to that of an equivalent random graph ($l\approx l_{\mathrm{rand}}$) but
with a much greater clustering coefficient
($C\gg C_{\mathrm{rand}}$)''~\cite{watts1999networks}.

\begin{figure}[ht!]
\center
\center{\includegraphics[width=1\linewidth]{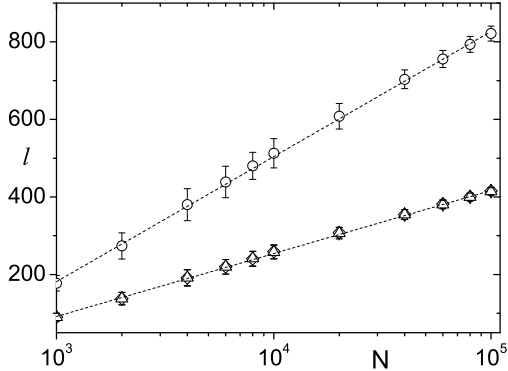}}\\
\caption{The average shortest path as a function of the number of nodes for
SW networks for $p=0.002$: circles are {\em A--k1},  triangles
are {\em A--k2},  diamonds are {\em R--k2}, and dashed lines
indicate fit functions.}
\label{pic:l(pN)}
\end{figure}

\begin{figure}[ht!]
\center{\includegraphics[width=1\linewidth]{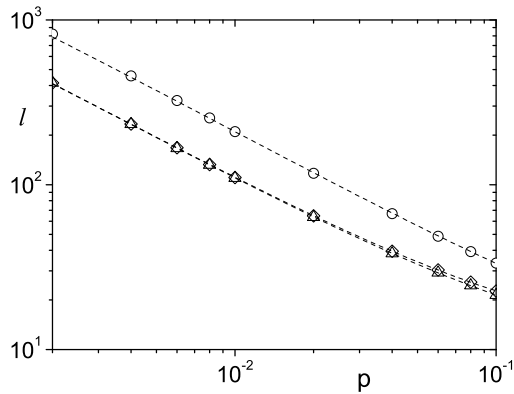}}
\caption{The average shortest path length as a function of the parameter
$p$ for systems of size $N=10^5$:  circles are {\em A--k1},
triangles are {\em A--k2}, diamonds are {\em R--k2}, and dashed
lines indicate fit functions of form (\ref{log-l-fit}) in all three cases. Error bars are of the symbol size.}
\label{l(p)plot(2)}
\end{figure}

To ensure that the constructed networks are indeed SW networks, we analyze
the dependence of the average shortest path length $l$ on the parameter $p$
and the system size $N$. We find that $l$ depends logarithmically on $N$ for
all $p>0$ for all networks (Fig.~\ref{pic:l(pN)}). Scaling relation
(\ref{lengthscaling}) is also observed in our data. We plot the average
shortest path as a function of the parameter $p$ for the network of size
$N=10^5$, and it is well approximated (see Fig.~\ref{l(p)plot(2)}) by
\begin{equation}
l=A \frac{\ln (pN)}{pk}+D.
\label{log-l-fit}
\end{equation}

We also calculate clustering coefficients in our models. For the network
{\em A--k1}, $C(p)\approx 0$. Strictly speaking, this model does not fully
satisfy the criteria for SW networks. For the networks {\em A--k2} and
{\em R--k2}, we plot $C(p)$ and compare the results with different
analytic estimates (Fig.~\ref{pic:clustering}). The agreement is good for
small $p$ ($p<0.01$), and the clustering coefficient for {\em A--k2} and
{\em R--k2} is close to $C(0)$, which equals $1/2$ for $k=2$.

\section{Small-world synchronization scheme}
\label{sec:SWmodel}

\subsection{Model of time evolution in the conservative algorithm}

The key property of the conservative synchronization scheme for PDES is the
preservation of causality. In the general case, causality is defined in
terms of the dependency matrix with elements $D(i,j)$, where $D(i,j)=1$ if
the process simulated by PE$_i$ depends on PE$_j$, and $D(i,j)=0$ otherwise.
Causality is preserved if the local virtual time (LVT) of PE$_i$ is lower
than LVT of those PE$_j$ on which PE$_i$ depends.

The time evolution begins with a flat profile $\tau_i(0)=0$, $i=1,2,\dots,N$.
To preserve causality, we randomly update those LVTs of PEs that are lower
than LVTs of the PEs on which they depend, i.e., using links defined by the
dependency matrix $D$. This leads to the rule
\begin{equation}
\tau_i(t+1)=\begin{cases}
\tau_i(t)+\eta_i&\text{if }\tau_i(t)\le\{\tau_{j}(t)\}_{D(i,j)=1},\\
\tau_i(t)&\text{otherwise},\end{cases}
\label{eq:update_scheme}
\end{equation}
where $ \eta_i$ is a random value drawn from the Poisson distribution,
$\{\tau_{j}(t)\}_{D(i,j)=1}$ is the set of all local times of the PEs
connected to PE$_i$ by local or long-range communication links,
$i=1,2,\dots,N$.

As is known, the model on the regular one-dimensional lattice belongs to the
KPZ universality class~\cite{korniss2000massively}. This can be seen by the
following reasoning. First, we represent Eq.~(\ref{eq:update_scheme}) in the
form
\begin{equation}
\tau_i(t+1)=\tau_i(t)+\Theta[\tau_{i-1}(t)-\tau_i(t)]
\Theta[\tau_{i+1}(t)-\tau_i(t)]\eta_i(t),
\label{eq:tau}
\end{equation}
neglecting long-range links and using the Heaviside step function $\Theta$.

Second, replacing differences between local times with the local slope
\begin{equation}
\phi_i=\tau_{i}-\tau_{i-1},
\label{eq:local_slope}
\end{equation}
we obtain the equation for the density of local minima (or the utilization):
\begin{equation}
u(t)=\frac{1}{N}\sum_{i=1}^N\Theta[-\phi_i(t)]\Theta[\phi_{i+1}(t)].
\label{eq:slopes}
\end{equation}

It was shown in~\cite{krug1990universal} that there is a finite-size
correction to the growth rate. The finite-size behavior of the average
profile speed is
\begin{equation}
\langle u(N)\rangle\simeq\langle u(\infty)\rangle+\frac{\text{const}}{N^{2(1-\alpha)}},
\label{eq:finite-size_u}
\end{equation}
where $\langle u(\infty)\rangle$ is the value of the average speed in the
asymptotic infinite number of PEs and $\alpha$ is the roughness exponent.
Equation~(\ref{eq:finite-size_u}) is confirmed by simulating LVT profile
growth. For the KPZ model $\langle u(\infty)\rangle=1/4$, while in the model
of evolution of the LVT profile, $\langle u(\infty)\rangle\approx0.24641$.
This is due to nonuniversal short-range correlations between the slopes in
the profile.

The average speed depends weakly on the type of distribution of the random
variable $\eta_i$. For $p=0$, it was shown in~\cite{shchur2004evolution}
that the average speed $\langle u\rangle_U=0.267(4)$ for a uniform
distribution of $\eta_i$, $\langle u\rangle_G=0.258(5)$ for a Gaussian
distribution of $\eta_i$, and $\langle u\rangle=0.246410(7)$ for a
Possion distribution of $\eta_i$.

It was argued by Korniss et al.~\cite{korniss2000massively} that the
coarse-grained slope $\hat\phi(x,\hat t)$ of the time horizon in the
continuum limit is evaluated according to the Burgers
equation~\cite{kardar1986dynamic}
\begin{equation}
\frac{\partial\hat\phi}{\partial\hat t}=\frac{\partial^2\hat\phi}{\partial x^2}-\lambda\frac{\partial\hat \phi^2}{\partial x}
\end{equation}
and the coarse-grained time profile $\hat\tau$, $\hat\phi=\partial\hat t/\partial x$
satisfies the KPZ equation
\begin{equation}
\frac{\partial\hat\tau}{\partial\hat t}=\frac{\partial^2\hat\tau}{\partial x^2}-\lambda\left(\frac{\partial\hat\tau}{\partial x}\right)^2,
\end{equation}
which should be extended with noise to capture the fluctuations.

We can expect that the evolution of the time profile belongs to the KPZ
universality class. Numerical analysis~\cite{korniss2000massively,shchur2004evolution}
supports this expectation. In the case of long-range links, we can expect
deviation from KPZ universality class.

In the case of long-range links, we can rewrite Eq.~(\ref{eq:tau})
as
\begin{equation}
\begin{split}
\tau_i(t+1)=\tau_i(t)+\Theta[\tau_{i-1}(t)-\tau_i(t)]\Theta[\tau_{i+1}(t)-\tau_i(t)]
\\
\prod_{\left\{D'(i,j)=1\right\}}\Theta[\tau_{j}(t)-\tau_i(t)]\eta_i(t),
\end{split}
\label{eq:tau-SW-k1}
\end{equation}
where the product is computed only for long-range links coming from the node
PE$_i$, which is denoted by the prime in $\left\{D'(i,j)=1\right\}$. The average time profile speed in this case is
\begin{equation}
\langle u(t)\rangle=\langle\Theta[-\phi_i(t)]\Theta[-\phi_{i+1}(t)]\prod_{\left\{D'(i,j)=1\right\}}\Theta[\tilde\phi_j(t)]\rangle,
\label{util-sw}
\end{equation}
where $\tilde\phi_j(t)=\tau_j-\tau_i$. It is clear from Eq.~(\ref{util-sw})
that additional dependencies decrease the LVT profile speed. In other words,
adding long-range links decreases the utilization. Simulations confirm this
observation.

\subsection{Simulations}

The simulation parameters are the number $N$ of PEs, the concentration $p$
of long-range links per PE, and the number $t$ of discrete simulation steps.
The matrix $D$ is randomly initialized with one of the construction
algorithms described in section~\ref{sec:SW}.

The average speed $\langle u \rangle$ and the average profile width $\langle w^2 \rangle$ are
calculated after each update using the respective expressions~(\ref{eq:u})
and (\ref{eq:w2}). For each set of parameters $N$ and $p$, we use 1500
different realizations of the random process running in parallel. The
parameter $p$ changes from $0.002$ to $0.1$, and the number $N$ of PEs
ranges from $10^3$ to $10^5$.

\subsubsection{Average speed}

\begin{figure}[ht!]
\center \begin{minipage}[h]{1\linewidth}
\center{\includegraphics[width=1\linewidth]{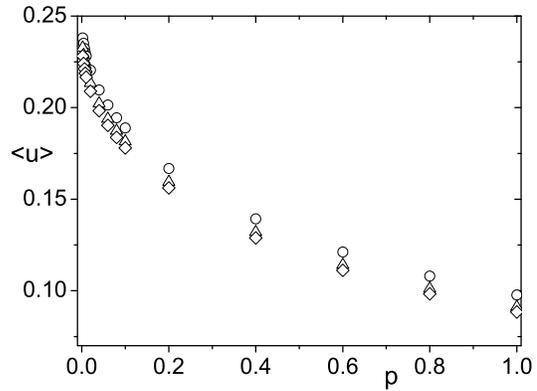}} a) {\em A--k1} \\
\end{minipage}
\vfill
\begin{minipage}[h]{1\linewidth}
\center{\includegraphics[width=1\linewidth]{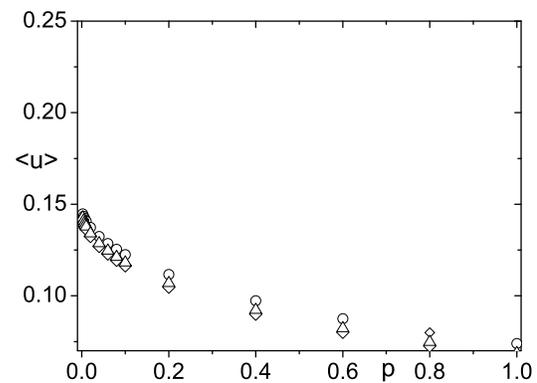}} \\b) {\em A--k2}
\end{minipage}
\vfill
\begin{minipage}[h]{1\linewidth}
\center{\includegraphics[width=1\linewidth]{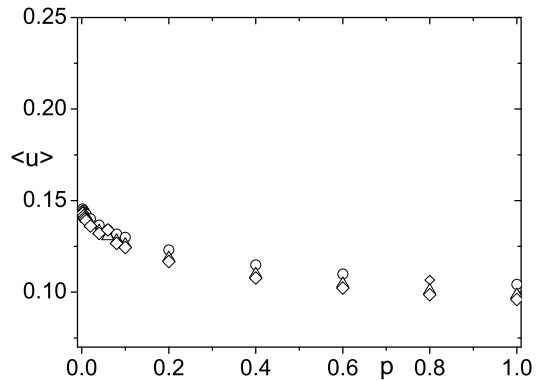}} c) {\em R--k2} \\
\end{minipage}
\vfill
\caption{The average speed $\langle u\rangle$ as a function of the
concentration $p$ of long-range links for different number of PEs:
circles for $N=10^3$,  triangles for $N=10^4$, and  diamonds for
$N=10^5$. Error bars are of the symbol size. The speed is averaged over time with the first 1000 time steps
omitted. }
\label{pic:u(p)_diff_networks}
\end{figure}

Figure~\ref{pic:u(p)_diff_networks} shows the dependence of the average
speed $\langle u\rangle$ on the concentration $p$ for three realizations of
the SW networks {\em A--k1}, {\em A--k2}, and {\em R--k2}: they
are the respective networks with two closest neighbors and $pN$ randomly
added links, with four closest neighbors and $pN$ randomly added links, and
with four closest neighbors and $pN$ randomly rewired links.

It can be seen in Fig.~\ref{pic:u(p)_diff_networks} that the average speed
$\langle u\rangle$ decreases as the concentration $p$ increases and is
smaller for the networks {\em A--k2}, and {\em R--k2} because of the
dependence on the next-to-neighbors. Strictly speaking, we should rewrite
Eqs.~(\ref{eq:tau-SW-k1}) and (\ref{util-sw}) in this case as
\begin{equation}
\begin{split}
\tau_i(t+1)=\tau_i(t)+\Theta[\tau_{i-1}(t)-\tau_i(t)]\Theta[\tau_{i+1}(t)-\tau_i(t)]\\
\Theta[\tau_{i-2}(t)-\tau_i(t)]\Theta[\tau_{i+2}(t)-\tau_i(t)]\\
\prod_{\left\{D'(i,j)=1\right\}}\Theta[\tau_{j}(t)-\tau_i(t)]\eta_i(t),
\end{split}
\end{equation}
Using Eq.~(\ref{eq:local_slope}), we obtain the expression for the average
profile speed on the network {\em A--k2}
\begin{equation}
\begin{split}
\langle u(t)\rangle=\langle\Theta[-\phi_i(t)]\Theta[\phi_{i+1}(t)]\\
\Theta[-\phi_{i-1}(t)-\phi_i(t)]\Theta[\phi_{i+2}(t)+\phi_{i+1}(t)]\\
\prod_{\left\{ D'(i,j)=1\right\}}\Theta[\tilde\phi_j(t)]\rangle,
\end{split}
\label{util-sw-k2}
\end{equation}
The presence of next-to-neighbors reduces the average speed
$\langle u\rangle$, and the average speed $\langle u_0\rangle=0.14674(7)$
for $p=0$. It can be seen that the speed remains positive for small
concentrations $p$, which means that the SW-synchronized simulation scheme
maintains a nonzero average utilization. For example, we have
$\langle u \rangle=0.221370(7)$ in {\em A--k1} for $p=0.01$ and
$\langle u_0\rangle=0.246410(7)$ for $p=0$.
It is seen from Figures~\ref{pic:u(p)_diff_networks}, that  the average speed $\langle u \rangle$
is hardly different for $N=10^4$ and $N=10^5$.

\begin{figure}[ht!]
\center{\includegraphics[width=1\linewidth]{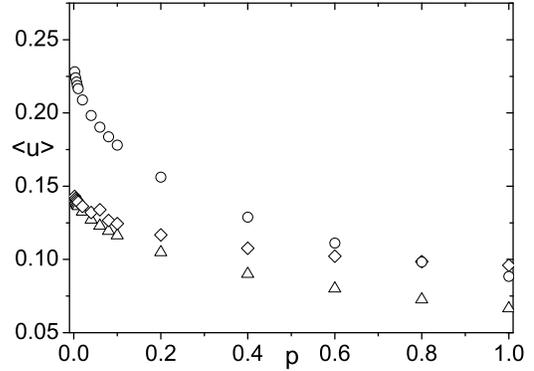}}
\caption{Comparison of the average speeds $\langle u\rangle$ on three
network realizations:  circles for {\em A--k1},  triangles for
{\em A--k2}, and  diamonds for {\em R--k2} for $N=10^5$. Error bars are of the symbol size.}
\label{pic:copmarison_of_u(p)}
\end{figure}

Figure~\ref{pic:copmarison_of_u(p)} shows the difference of the average
speed depending on the SW network realization in the systems of $N=10^5$ PEs. For small values of the
parameter $p$, the difference between the average speed on the network
{\em A--k1} and networks {\em A--k2} and {\em R--k2} is
significant. This is expected from Eqs.~(\ref{util-sw}) and
(\ref{util-sw-k2}). In the latter equation, the additional terms slow the
interface growth speed.

For $p$ close to unity, the average speeds on the networks {\em A--k1}
and {\em R--k2} are approximately the same. This can be explained by
comparing the average amount of dependencies in these networks. For $p=1$,
the network {\em A--k1} has $N(1{+}p){=}2N$ links between the PEs, and
the network {\em R--k2} also has $2N$ links. We can conclude that the
average speed of the LVT profile on SW networks mainly depends on the number
of links in the communication network.

\begin{figure}[ht!]
\center{\includegraphics[width=1\linewidth]{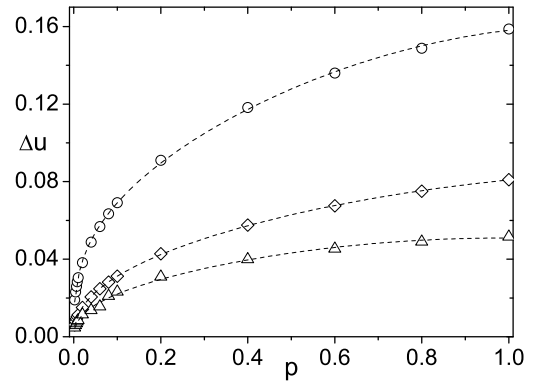}}
\caption{The difference between the speeds $\Delta u=\langle u\rangle-
\langle u_{0}\rangle $ as a function of the concentration $p$ of long-range
links for $N=10^5$ for three network realizations:  circles for
{\em A--k1},  triangles for {\em A--k2},  diamonds for
{\em R--k2}, and dashed lines for fit~(\ref{eq:deltaU}). Error bars are of the symbol size.}
\label{pic:all_delta_u_lin}
\end{figure}

The dependence of the average speed $\langle u\rangle$ on the parameter $p$
is nonlinear. Let $\Delta u$ be the difference between the average speed
$\langle u\rangle$ on a SW network and the average speed $\langle u_0\rangle$
on a regular lattice:
$$
\Delta u=\langle u_0\rangle-\langle u\rangle.
$$
The difference $\Delta u$ between the speeds is well approximated by a
power-law function (Fig.~\ref{pic:all_delta_u_lin}):
\begin{equation}
\Delta u(p,N)\sim p^{B(N)}.
\label{eq:deltaU}
\end{equation}

\begin{figure}[ht!]
\center{\includegraphics[width=1\linewidth]{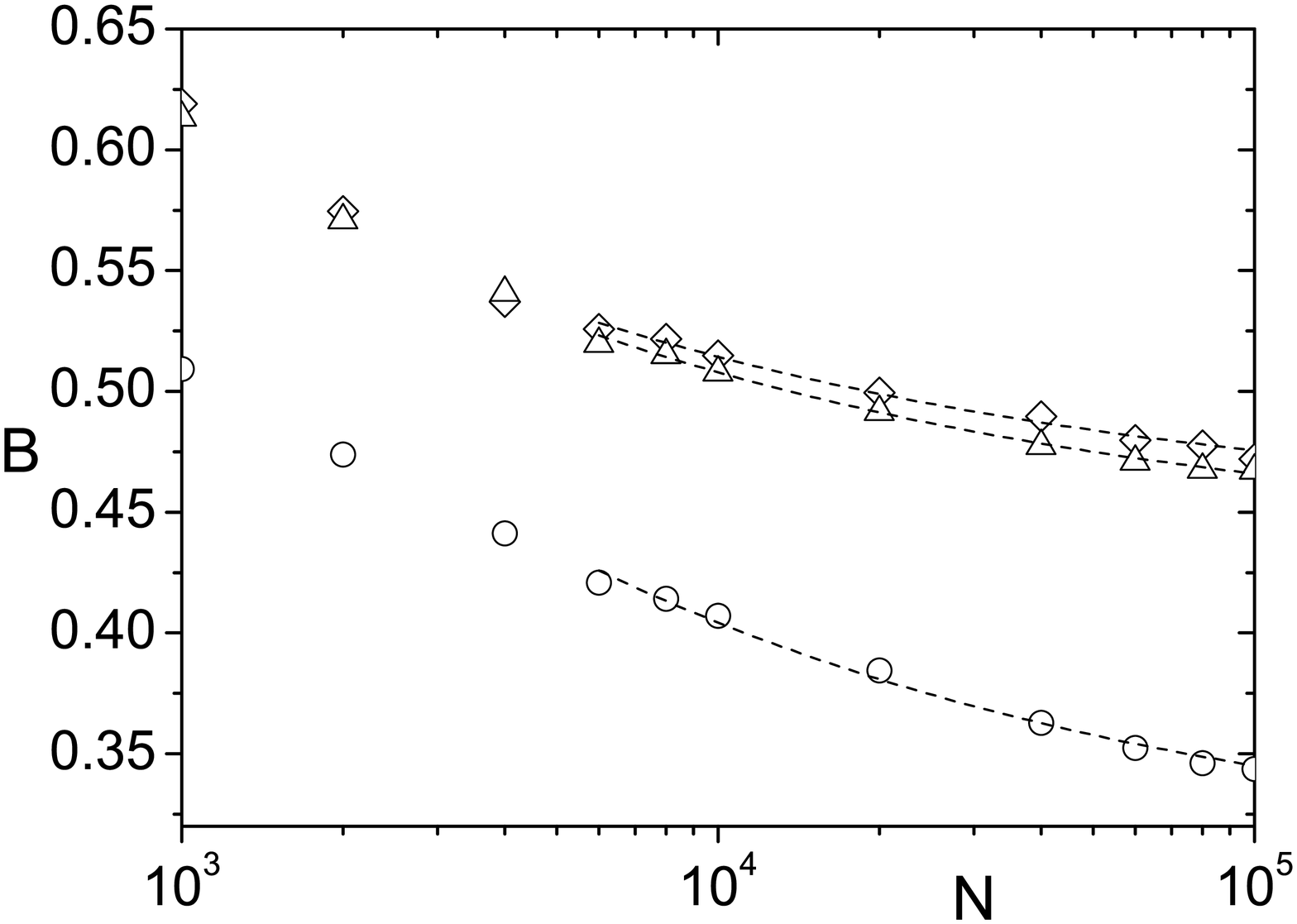}}
\caption{The exponent $B$ as a function of the number $N$ of PEs for three
network realizations:  circles for {\em A--k1},  triangles for
{\em A--k2},  diamonds for {\em R--k2}, and dashed lines for
fits (see discussion in the text). Error bars are of the symbol size.}
\label{pic:B(N)}
\end{figure}

The values of the exponent $B(N)$ are given in Table~\ref{tab:deltaU}. It
can be seen from the table that the exponent $B(N)$ decreases with the
number of PEs. Figure~\ref{pic:B(N)} shows the exponent $B(N)$ for three SW
network realizations. We find the asymptotic behavior of $B(N)$ in the limit
of a large number of PEs by approximating with the function
\begin{equation}
B(N)\approx B+A\frac{\ln N}{\sqrt{N}}.
\label{eq:B(N)}
\end{equation}

In the limit as $N\to\infty$, the exponent $B(N)$ approaches the values
$B=0.306(4)$ for the network {\em A--k1}, $B=0.439(2)$ for the network
{\em A--k2}, and $B= 0.450(2)$ for the network {\em R--k2}.

The behavior of the average speed in all three cases is not universal. The
exponents $B$ in the last two networks are very close to each other but
differ from the exponent $B$ in the network {\em A--k1}. This is
probably due to the topological differences between the networks. The most
significant topological difference between the network {\em A--k1} and
the networks {\em A--k2} and {\em R--k2} is the presence of
clustering. The network {\em A--k1} has a zero clustering coefficient,
while the other two networks are highly clustered. We can also conclude that
the particular way the SW topology is constructed (by either adding or
rewiring links) does not play an important role. More investigations should be done
with more networks of different topologies for the detailed classification of system behaviour.

\begin{table}[h!]
\begin{center}
\caption{Exponent $B$ (see Eq.(\ref{eq:deltaU})) for three realizations of
the SW networks and for different numbers $N$ of PEs. }
\label{tab:deltaU}
\begin{tabular}{|c||c|c|c|}
\hline
N & {\em A--k1} & {\em A--k2} & {\em R--k2} \\ \hline
$10^3$ & 0.509(2) & 0.613(4) & 0.62(1) \\ \hline
$10^4$ & 0.407(4) & 0.508(1) & 0.515(2) \\ \hline
$10^5$ & 0.344(7) & 0.467(4) & 0.472(8) \\ \hline
$\infty$ & 0.306(4) & 0.439(2) & 0.450(2) \\ \hline
\end{tabular}
\end{center}
\end{table}


\subsubsection{Average profile width}

\begin{figure}[ht!]
\center \begin{minipage}[h]{1\linewidth}
\center{\includegraphics[width=1\linewidth]{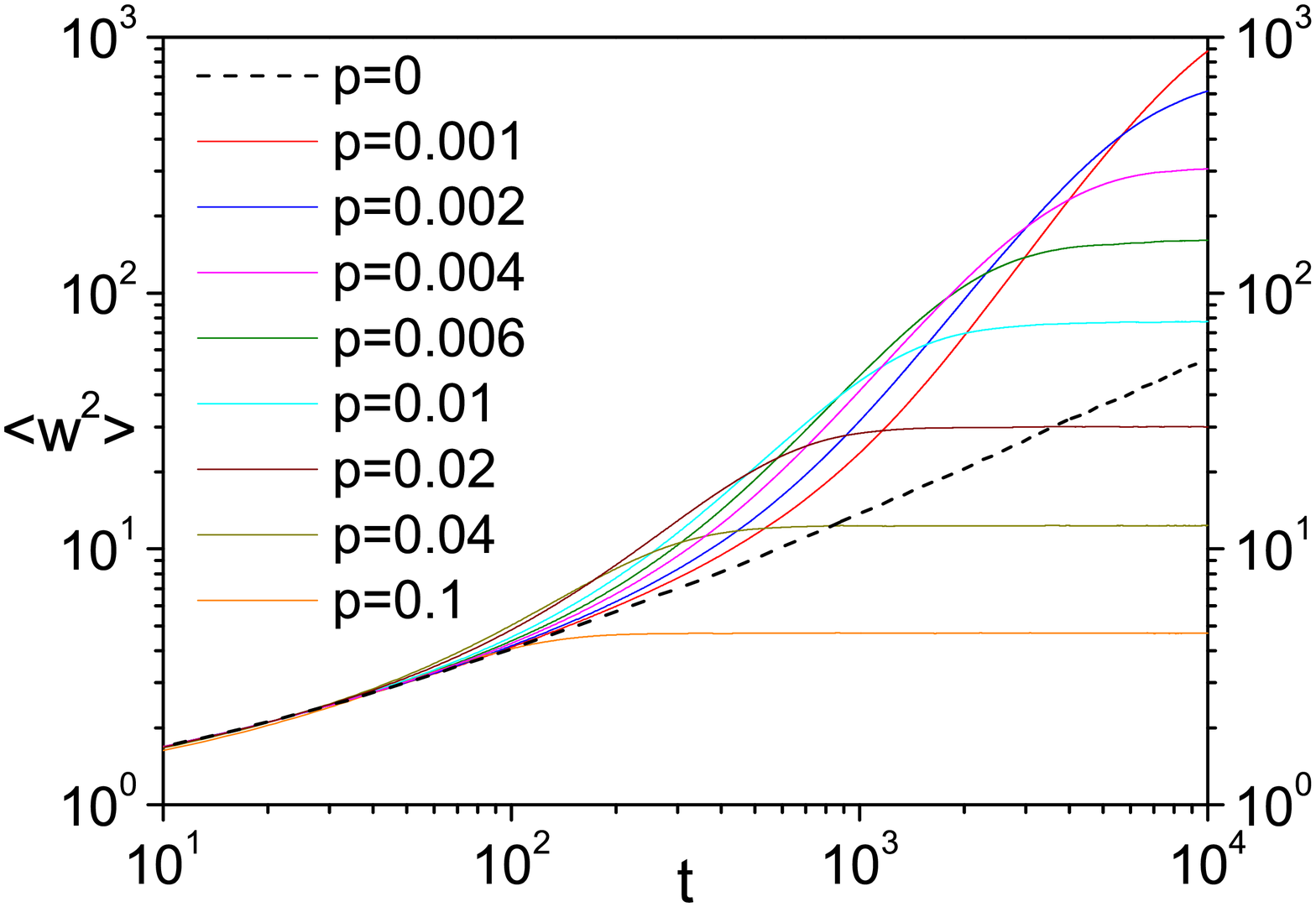}} a) {\em A--k1} \\
\end{minipage}
\vfill
\begin{minipage}[h]{1\linewidth}
\center{\includegraphics[width=1\linewidth]{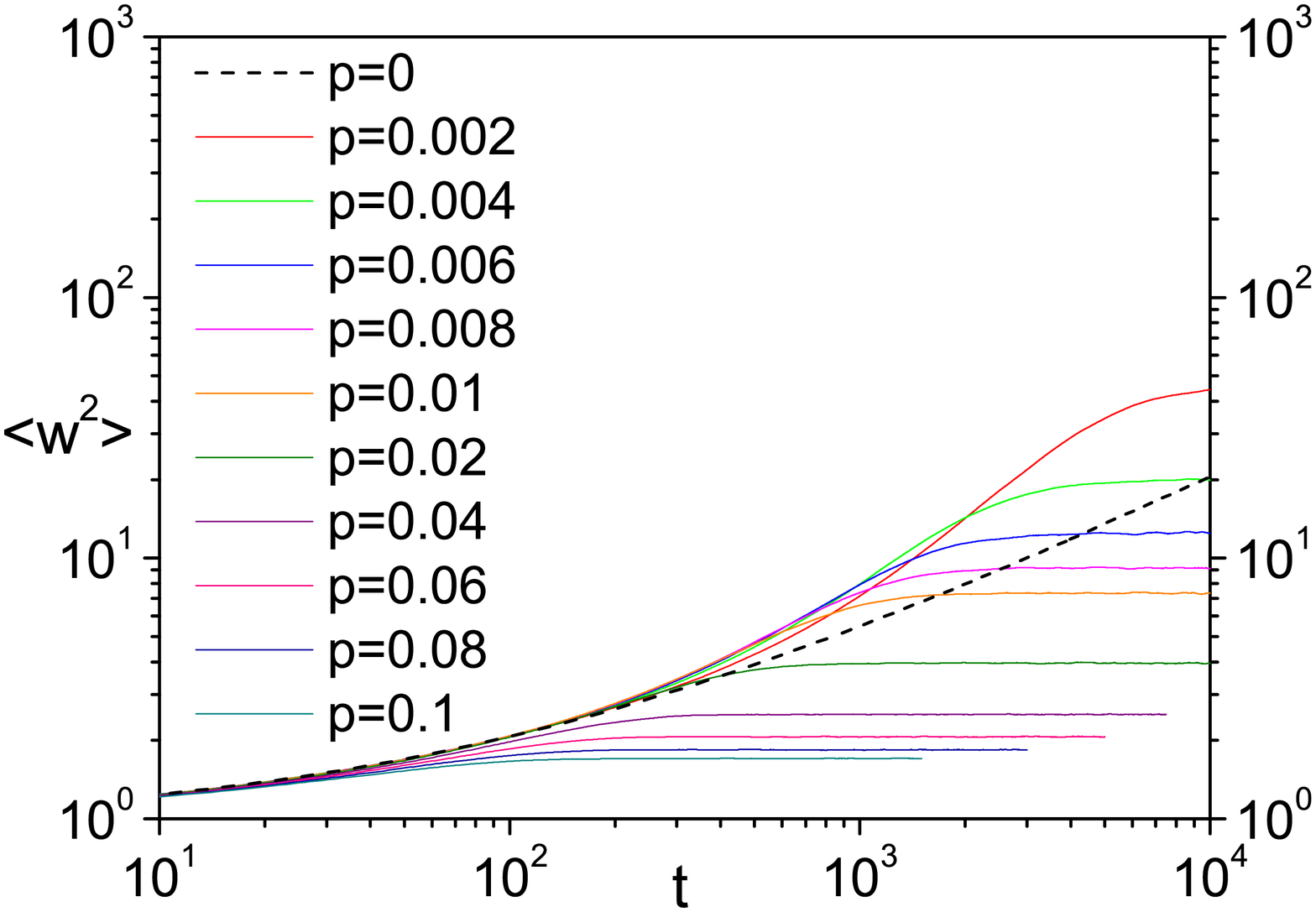}} \\b) {\em A--k2}
\end{minipage}
\vfill
\begin{minipage}[h]{1\linewidth}
\center{\includegraphics[width=1\linewidth]{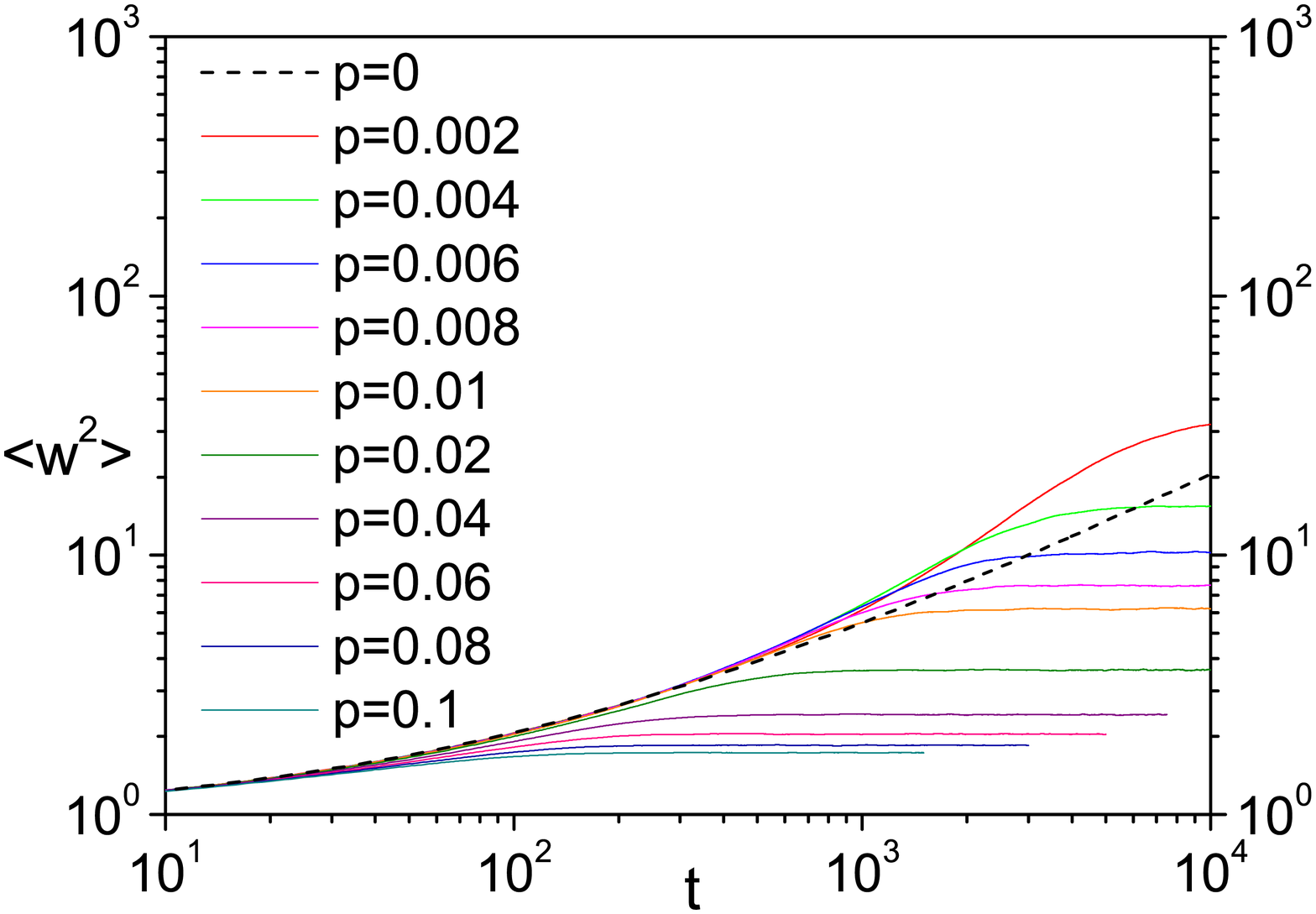}} c) {\em R--k2}\\
\end{minipage}
\vfill
\caption{(Color online) The average width $\langle w^2\rangle$ as a function of time for
the system size $N=10^4$ and different values of the parameter $p$: the
average is taken over 1500 independent runs. The black dotted line
corresponds to $p=0$ (KPZ universality class), and the solid lines
correspond to different values of parameter $p>0$. The order of solid lines from top to bottom corresponds to the figure legend.}
\label{pic:w2(t)}
\end{figure}

Figure~\ref{pic:w2(t)} shows the time dependence of the average width
$\langle w^2\rangle$ for three SW network realizations with $N=10^4$ PEs. It
can be seen that the profile width grows exponentially with time,
\begin{equation}
\langle w^2(t)\rangle\sim t^{2\beta},
\end{equation}
and saturates after a time $t_\times$. The larger the value of $p$ is, the slower
the width grows, and the lower the saturation value $\langle w^2_{\infty}\rangle$
is. The width saturates much earlier in the presence of long-range links
than in the case $p=0$. The width saturates after a sufficiently large time
$t_\times\approx10^6$ on a regular lattice of size $N=10^4$~\cite{korniss2000massively}
and after a time $t_\times<10^4$ on SW networks, even for a very small
concentration $p$.

\begin{figure}[ht!]
\center \begin{minipage}[h]{1\linewidth}
\center{\includegraphics[width=1\linewidth]{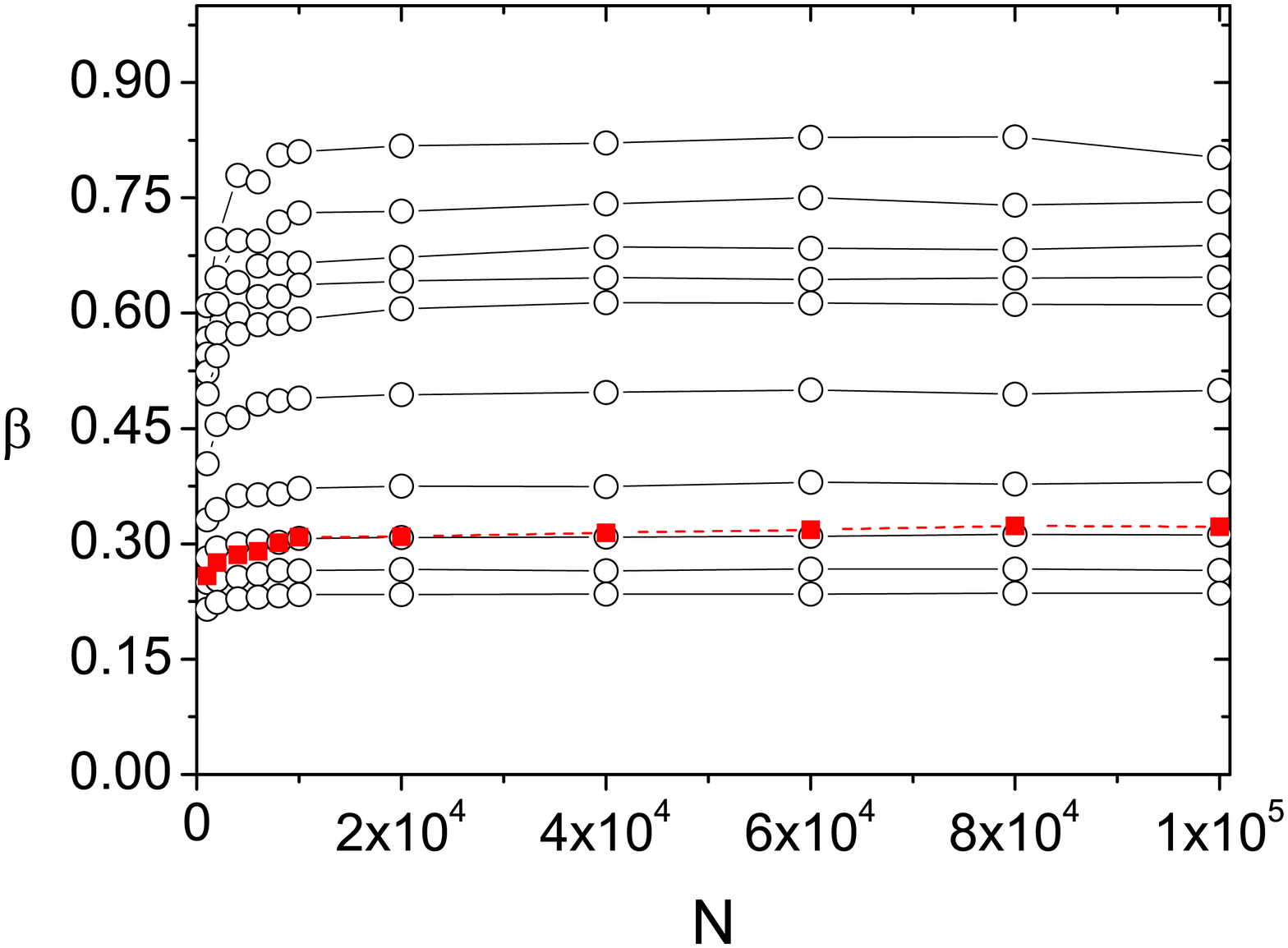}} a) {\em A--k1} \\
\end{minipage}
\vfill
\begin{minipage}[h]{1\linewidth}
\center{\includegraphics[width=1\linewidth]{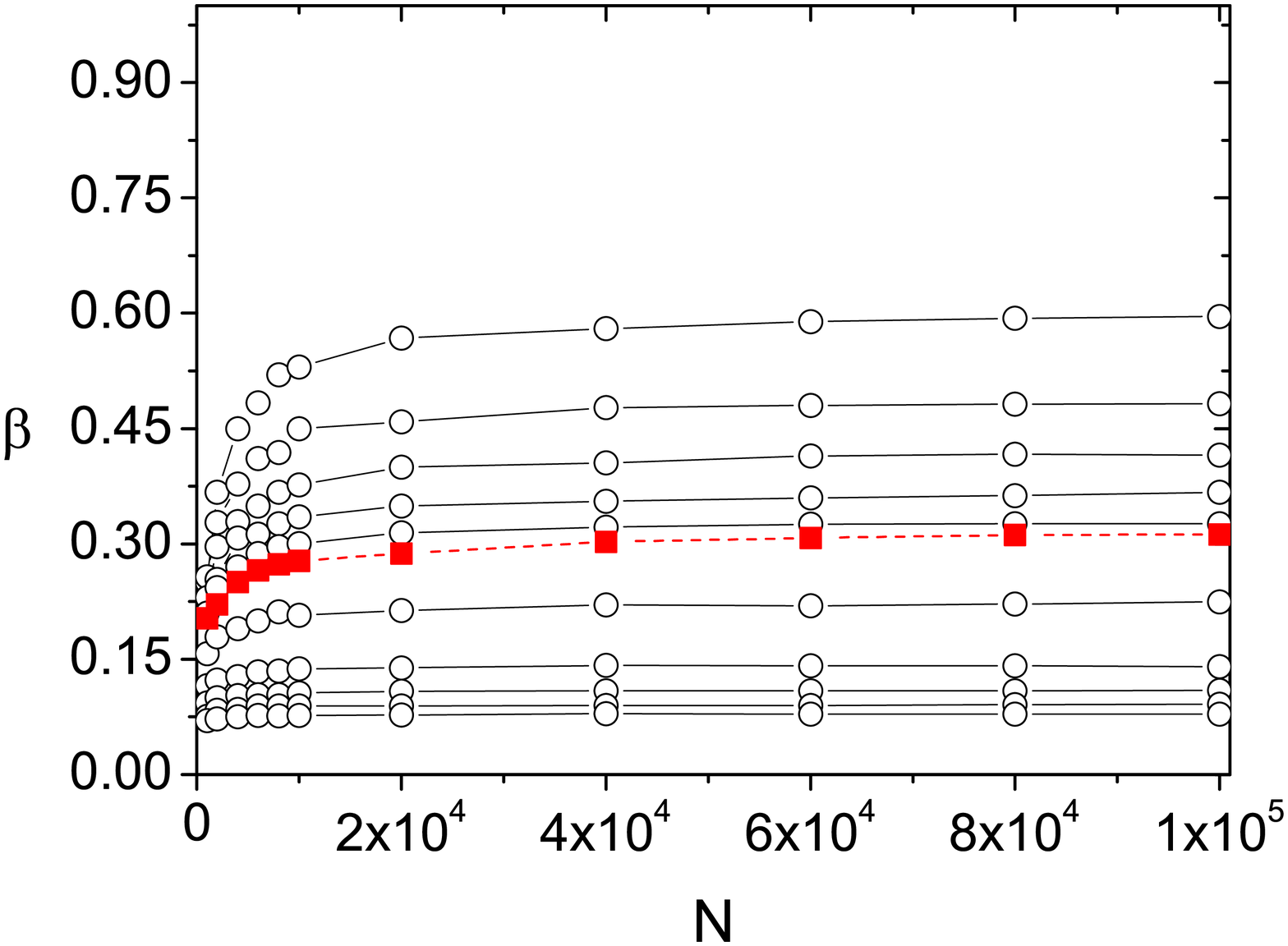}} \\b) {\em A--k2}
\end{minipage}
\vfill
\begin{minipage}[h]{1\linewidth}
\center{\includegraphics[width=1\linewidth]{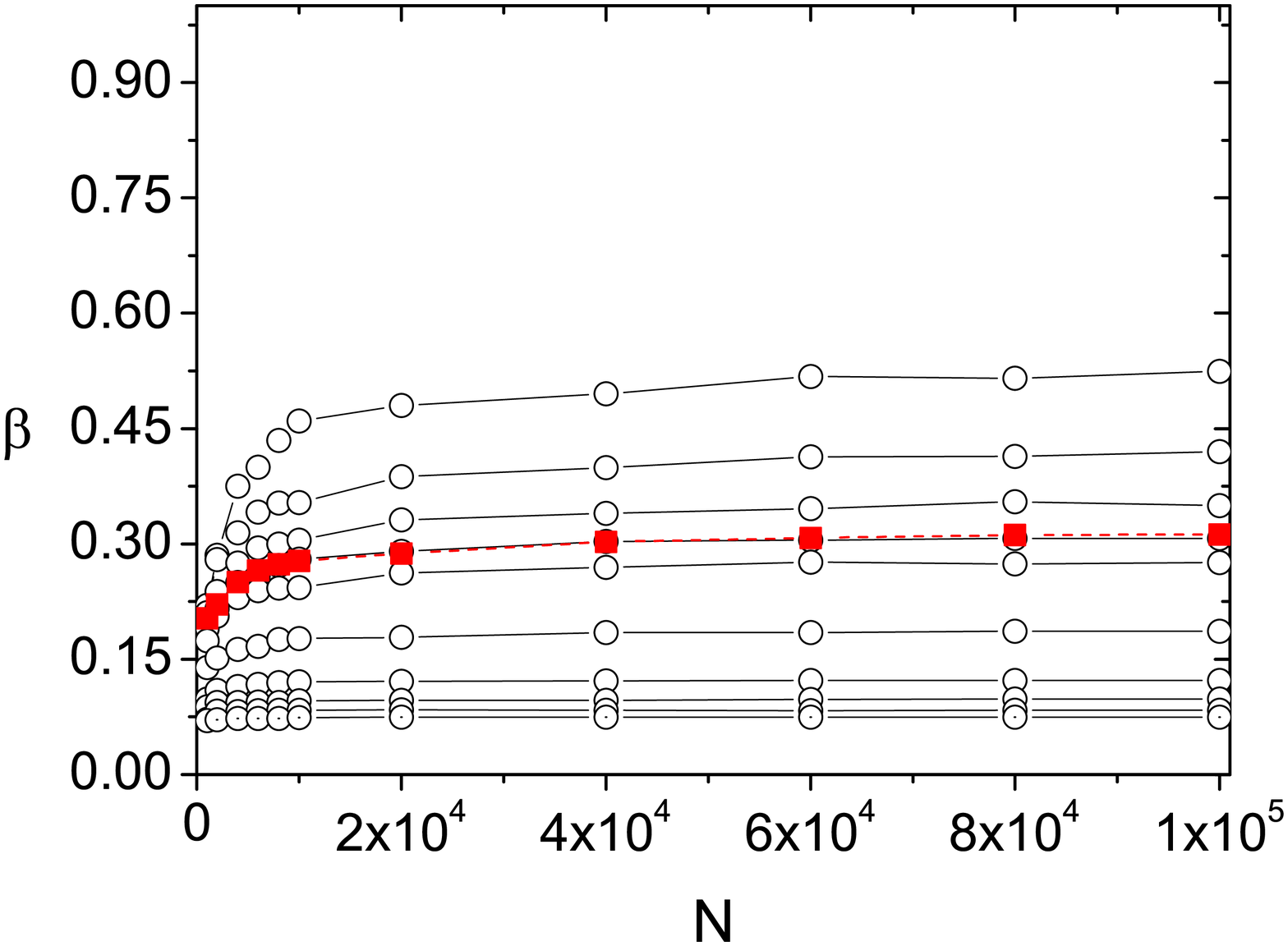}} c) {\em R--k2} \\
\end{minipage}
\vfill
\caption{(Color online) The growth exponent $\beta$ as function of the system size $N$: the values of $p$ change from top to bottom: 0.002, 0.004, 0.006, 0.008, 0.01,
0.02, 0.04, 0.06, 0.08, 0.1; the dashed line with solid squares corresponds to the regular network
with $p=0$. Error bars are of the symbol size.}
\label{pic:beta(N)}
\end{figure}

It can also be seen from Fig.~\ref{pic:beta(N)} that the growth exponent
$\beta$ for systems with sufficiently many PEs ($N>2\cdot10^3$) becomes
constant and independent of $N$. We find the asymptotic values of $\beta$ as
$N\to\infty$ using an approximation with power-law corrections. The values
of $\beta$ for systems on three SW realizations and various values of the
parameter $p$ are listed in Table~\ref{tab:beta(p)}. Clearly, the growth
exponent $\beta$ decreases as the concentration $p$ increases.

\begin{figure}[ht!]
\center{\includegraphics[width=1\linewidth]{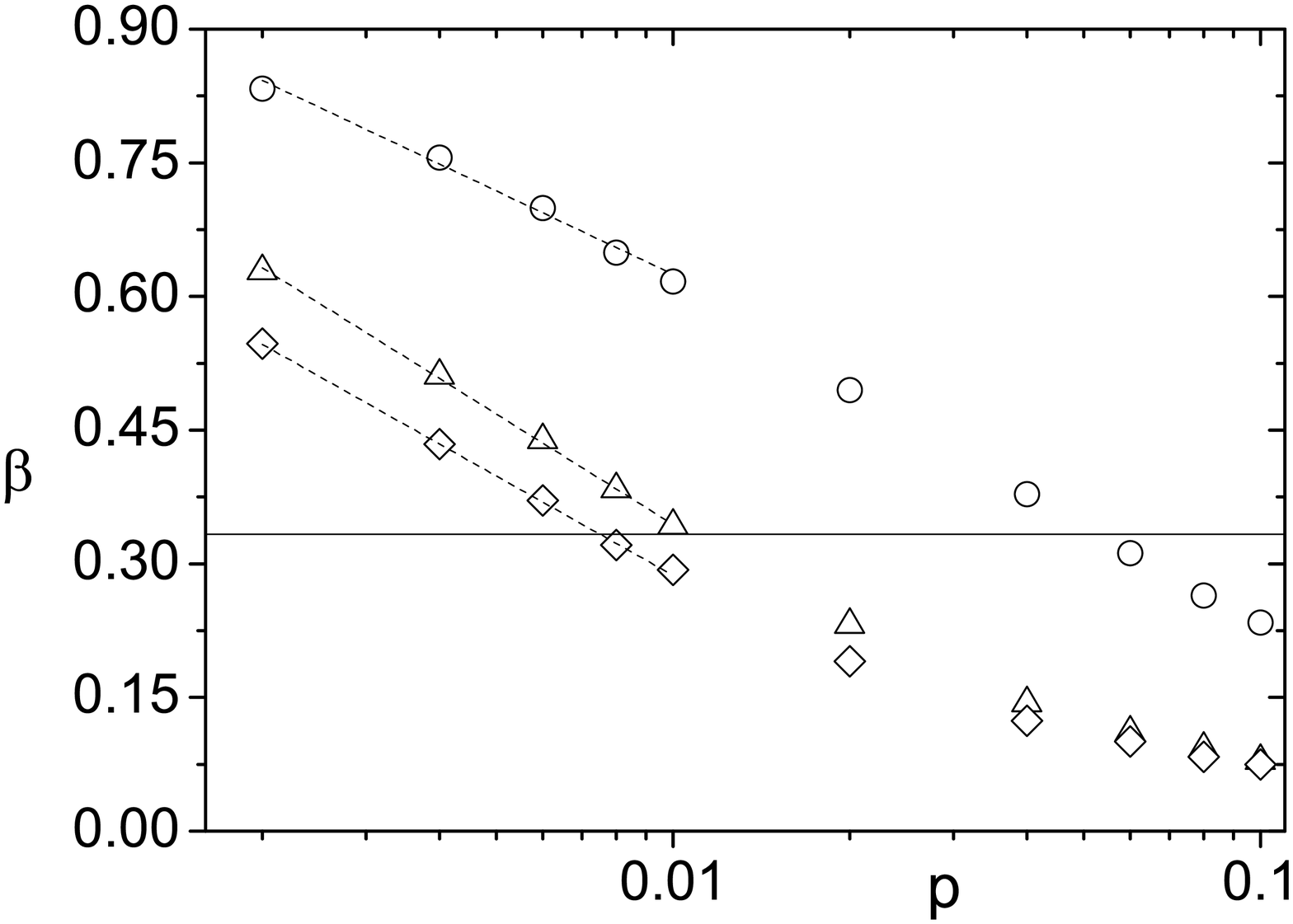}}
\caption{The exponent $\beta$ as a function of the concentration $p$: the
solid line shows the value $\beta=1/3$, the dashed lines are results
of the fitting, circles correspond to {\em A--k1} with the fit
$\beta\sim-0.311(2)\ln(p)$,  triangles correspond to {\em A--k2} with
the fit $\beta\sim-0.179(2)\ln(p)$, and diamonds correspond to
{\em R--k2} with the fit $\beta\sim-0.161(3)\ln(p)$. Error bars are of the symbol size.}
\label{pic:beta(p)}
\end{figure}

Figure~\ref{pic:beta(p)} shows the exponent $\beta$ as a function of the
parameter $p$. We find that for $p>0$, the exponent $\beta$ depends
logarithmically on the concentration $p$:
\begin{equation}
\beta(p)\sim-\ln(p)
\end{equation}

It is important that change of the exponent $\beta$ from SW lattices to
regular lattices is singular, as can be seen from Table~\ref{tab:beta(p)}.
Even a very small value of $p$ changes the exponent dependence from $1/3$
to $-\ln(p)$.

\begin{table}[ht]
\begin{center}
\caption{Dependence of the exponent $\beta$ on the concentration $p$ of
long-range links. }
\label{tab:beta(p)}
\begin{tabular}{|l||l|l|l|}
\hline
p& {\em A--k1}& {\em A--k2}& {\em R--k2} \\ \hline
0 & 0.33280(4) & 0.333(4) & $0.333(4)$ \\ \hline
\hline
0.002 & 0.833(3) & 0.629(6) & $0.547(6)$ \\ \hline
0.004 & 0.756(3) & 0.511(5) & $0.43(2)$ \\ \hline
0.006 & 0.699(2) & 0.439(8) & $0.371(9)$ \\ \hline
0.008 & 0.649(2) & 0.38(1) & $0.321(3)$ \\ \hline
0.01 & 0.617(4) & 0.343(4) & $0.293(6)$ \\ \hline
0.02 & 0.4949(8) & 0.232(3) & $0.191(4)$ \\ \hline
0.04 & 0.3783(7) & 0.144(1) & $0.124(2)$ \\ \hline
0.06 & 0.312(1) & 0.1106(8) & $0.101(2)$ \\ \hline
0.08 & 0.2646(8) & 0.093(1) & $0.0836(3)$ \\ \hline
0.1 & 0.2341(3) & 0.0791(5) & $0.0745(2)$ \\ \hline
\end{tabular}
\end{center}
\end{table}

\begin{figure}[ht!]
\center \begin{minipage}[h]{1\linewidth}
\center{\includegraphics[width=1\linewidth]{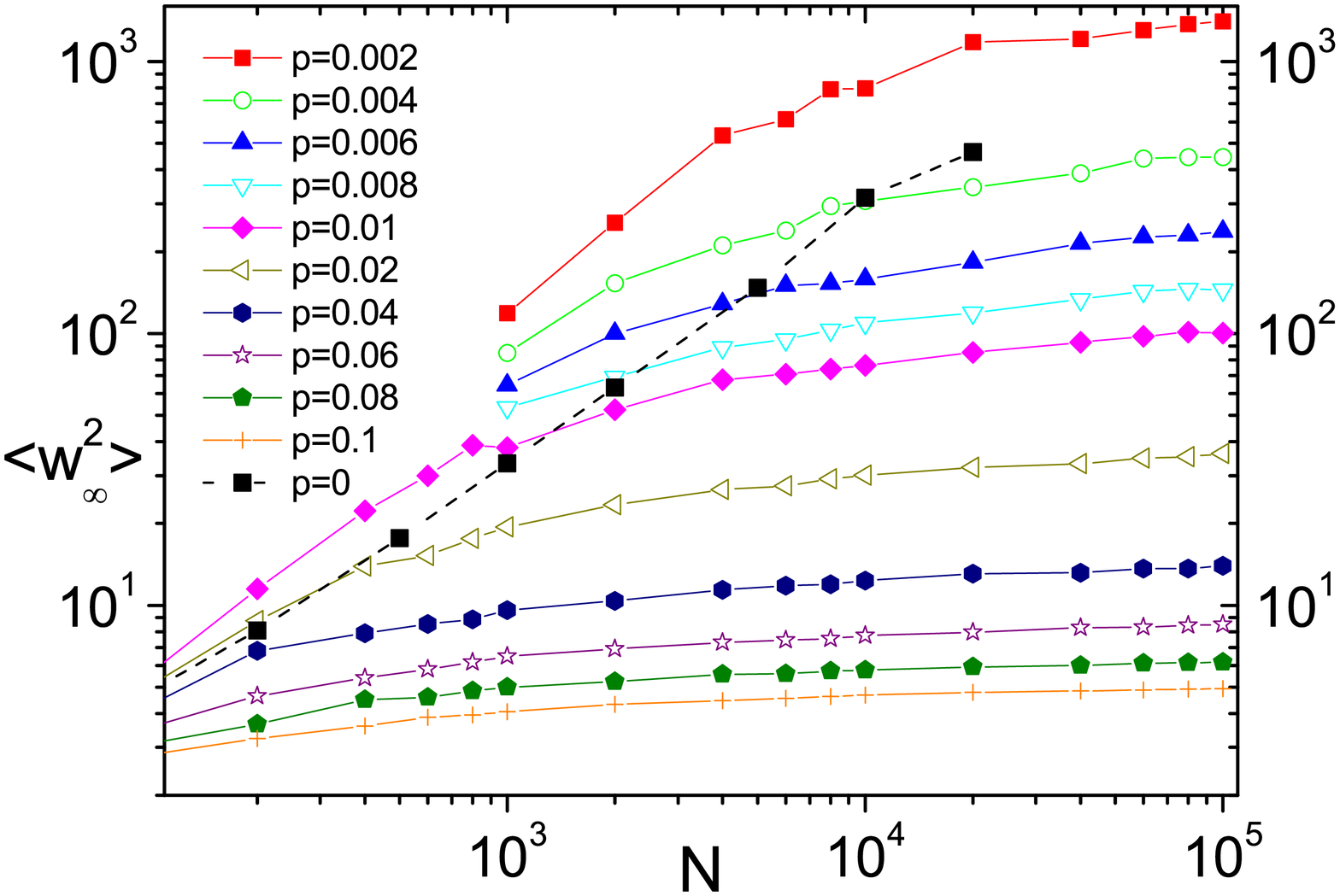}} a) {\em A--k1} \\
\end{minipage}
\vfill
\begin{minipage}[h]{1\linewidth}
\center{\includegraphics[width=1\linewidth]{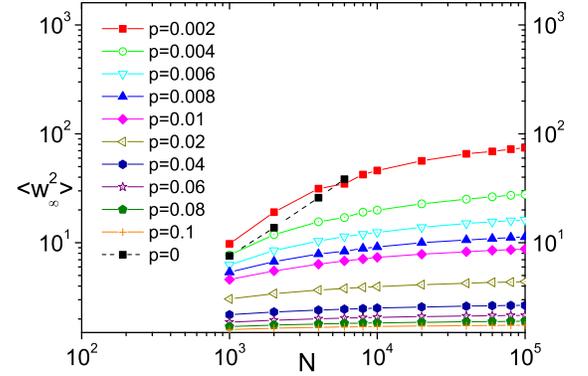}} \\b) {\em A--k2}
\end{minipage}
\vfill
\begin{minipage}[h]{1\linewidth}
\center{\includegraphics[width=1\linewidth]{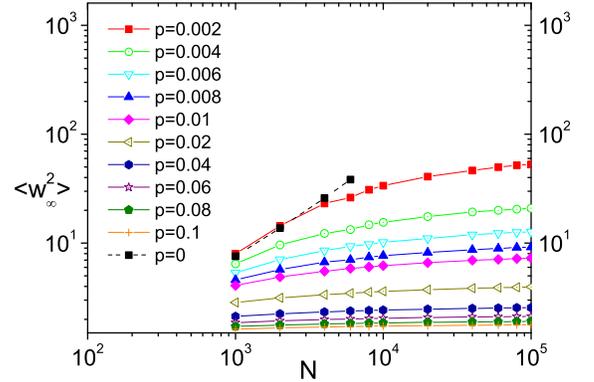}} c) {\em R--k2} \\
\end{minipage}
\vfill
\caption{(Color online) The average steady-state width as a function of the number $N$ of PEs. Dashed line corresponds to $p=0$. The order of solid lines from top to bottom corresponds to the figure legend. }
\label{pic:w2(N)}
\end{figure}

For each set of parameters $N$ and $p$, we measure the saturation value
$\langle w^2_{\infty}(N,p)\rangle$ of the width by averaging the width over
times $t\gg t_x$. Figure~\ref{pic:w2(N)} shows the steady-state width
$\langle w^2_{\infty}\rangle$ as a function of the number $N$ of PEs for
different concentrations $p$. In the case $p=0$, the steady-state width
scales as
\begin{equation}
\langle w^2_{\infty}\rangle\sim N^{2\alpha}
\label{eq:w2KPZ}
\end{equation}
where $\alpha$ is the roughness exponent approximately equals to $1/2$ (KPZ universality
class).

In contrast to expression~(\ref{eq:w2KPZ}), the average width on SW networks
does not increase with the number of PEs$, \langle w^2_{\infty}(N)\rangle=
\mathrm{const}$, i.e., the roughness exponent $\alpha=0$. The asymptotic
values of $\langle w^2_{\infty}\rangle$ in the limit of infinitely many PEs
for all three SW network realizations and different values of the parameter
$p$ are shown in Table~III. 
The average LVT profile width $\langle w^2_{\infty}\rangle$ decreases as
the parameter $p$ increases. Therefore, desynchronization is finite, and
its value decreases as $p$ increases.

\begin{table}[ht]
\begin{center}
\caption{Steady-state width $\langle w^2_{\infty}\rangle$ for various SW
network realizations.}
\label{tab:w2(p)}
\begin{tabular}{|l||c|c|c|}
\hline
p& {\em A--k1}& {\em A--k2}& {\em R--k2} \\ \hline
0 & $\sim N$ & $\sim N $ & $\sim N $ \\ \hline
\hline
0.002 & $2401\pm 687$ & 107(5) & 76(3)\\ \hline
0.004 & $1092 \pm 295$ & 40(3) & 25.1(8)\\ \hline
0.006 & $537\pm203$ & 20.8(6) & 15.3(7)\\ \hline
0.008 & $273\pm45$ & 13.3(3) & 11.1(4)\\ \hline
0.01 & $151\pm5$ & 10.7(7) & 8.3(1)\\ \hline
0.02 & $46\pm3$ & 4.88(8) & 4.26(6)\\ \hline
0.04 & $16.7\pm0.7$ & 2.81(1) & 2.69(2)\\ \hline
0.06 & $9.4\pm0.2$ & 2.22(1) & 2.201(8)\\ \hline
0.08 & $6.70\pm 0.07$ & 1.951(4) & 1.943(6)\\ \hline
0.1 & $5.18\pm 0.05$ & 1.783(5) & 1.827(8)\\ \hline
\end{tabular}
\end{center}
\end{table}

It can also be seen from Fig.~\ref{pic:w2(N)} that the saturation value
$\langle w^2_{\infty}\rangle$ is one order of magnitude less on the SW
networks {\em A--k2} and {\em R--k2} than on the network
{\em A--k1}. The reason is that additional dependencies in the system
cause additional synchronization between PEs.

\begin{figure}[ht!]
\center \begin{minipage}[h]{1\linewidth}
\center{\includegraphics[width=1\linewidth]{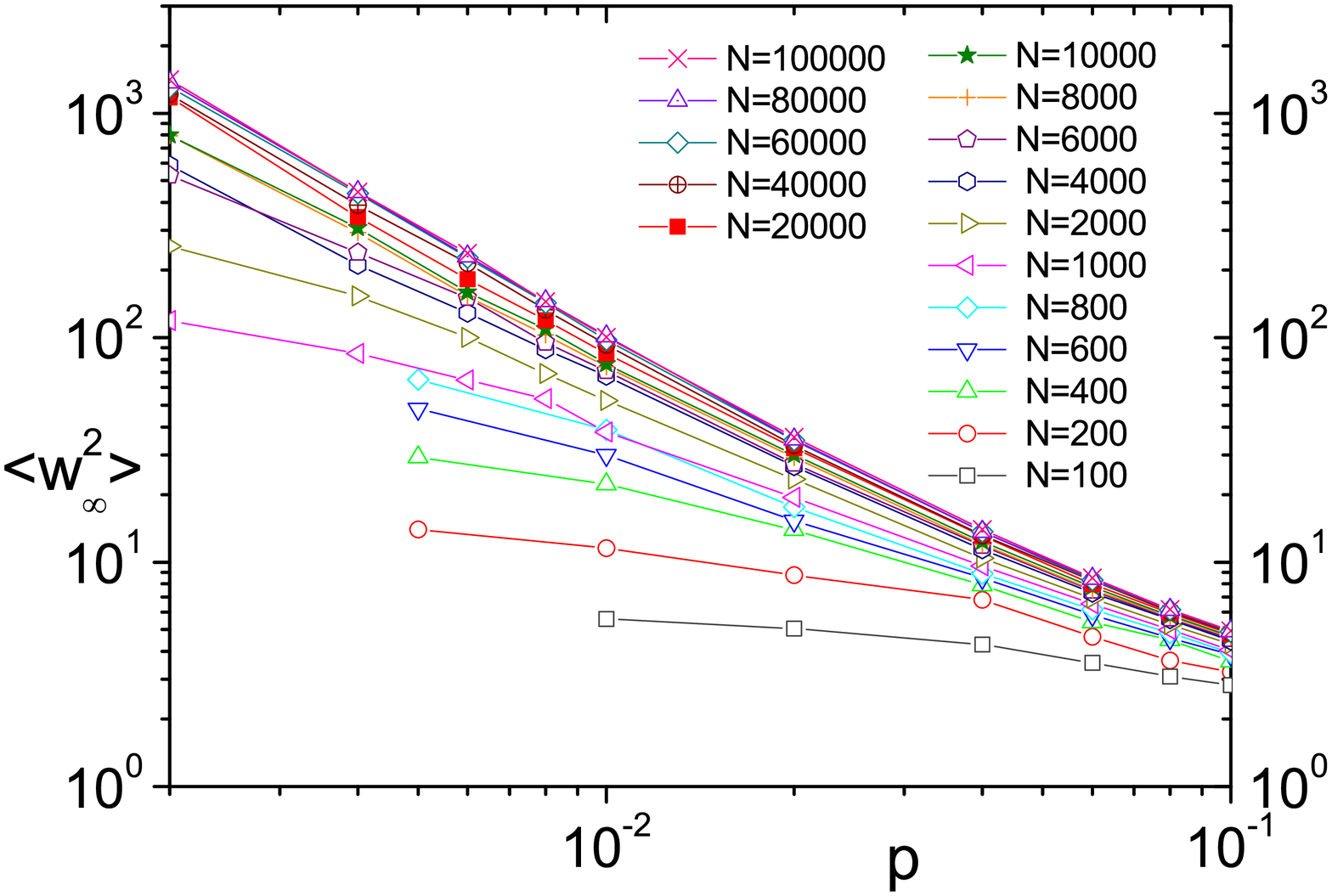}} a) {\em A--k1} \\
\end{minipage}
\vfill
\begin{minipage}[h]{1\linewidth}
\center{\includegraphics[width=1\linewidth]{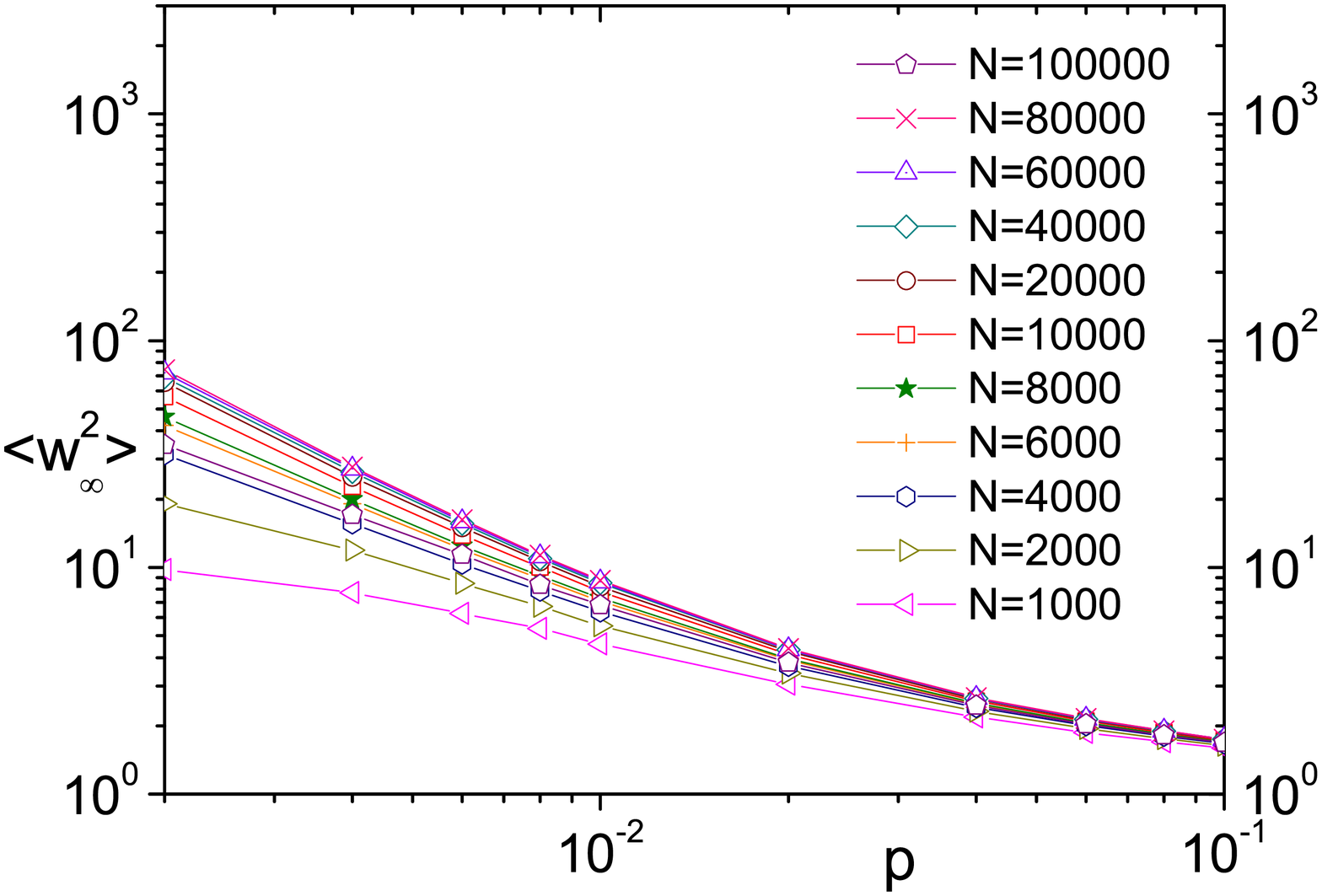}} \\b) {\em A--k2}
\end{minipage}
\vfill
\begin{minipage}[h]{1\linewidth}
\center{\includegraphics[width=1\linewidth]{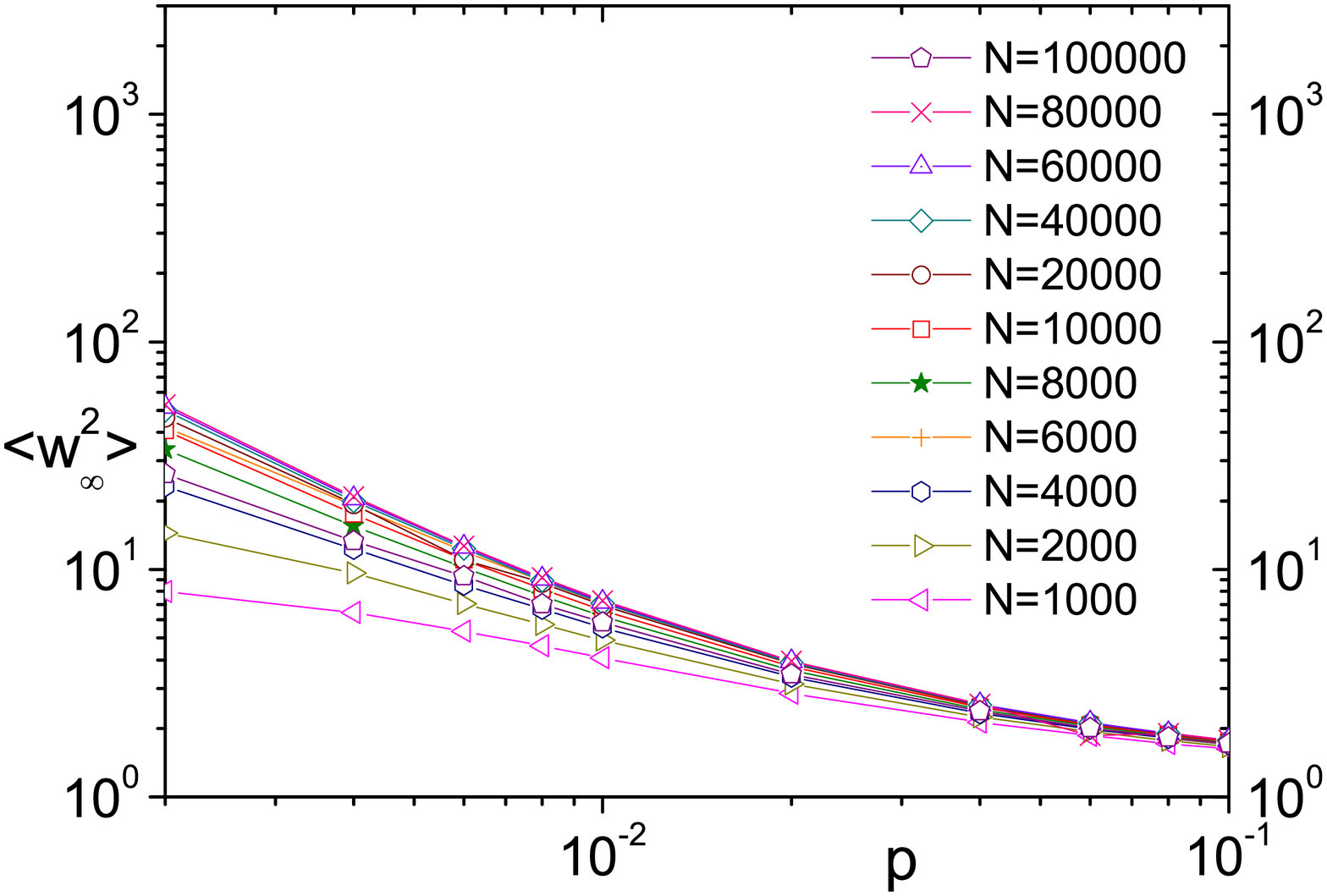}} c) {\em R--k2} \\
\end{minipage}
\vfill
\caption{(Color online) The average steady-state width $\langle w^2_{\infty}\rangle$ as a function of the parameter $p$. The order of lines from top to bottom corresponds to the figure legend. }
\label{pic:w2(p)}
\end{figure}

For large $p$, the average LVT profile width $\langle w^2_{\infty}\rangle$
on systems with various numbers of PEs has approximately the same small
value (Fig.~\ref{pic:w2(p)}). A small average width indicates that PEs are
well synchronized, but the utilization (average speed) is low in this case.
This indicates that there is some tradeoff between synchronization and
utilization, and a compromise can be achieved with a suitable rearrangement
of the communication network.

\begin{figure}[ht!]
\center \begin{minipage}[h]{1\linewidth}
\center{\includegraphics[width=1\linewidth]{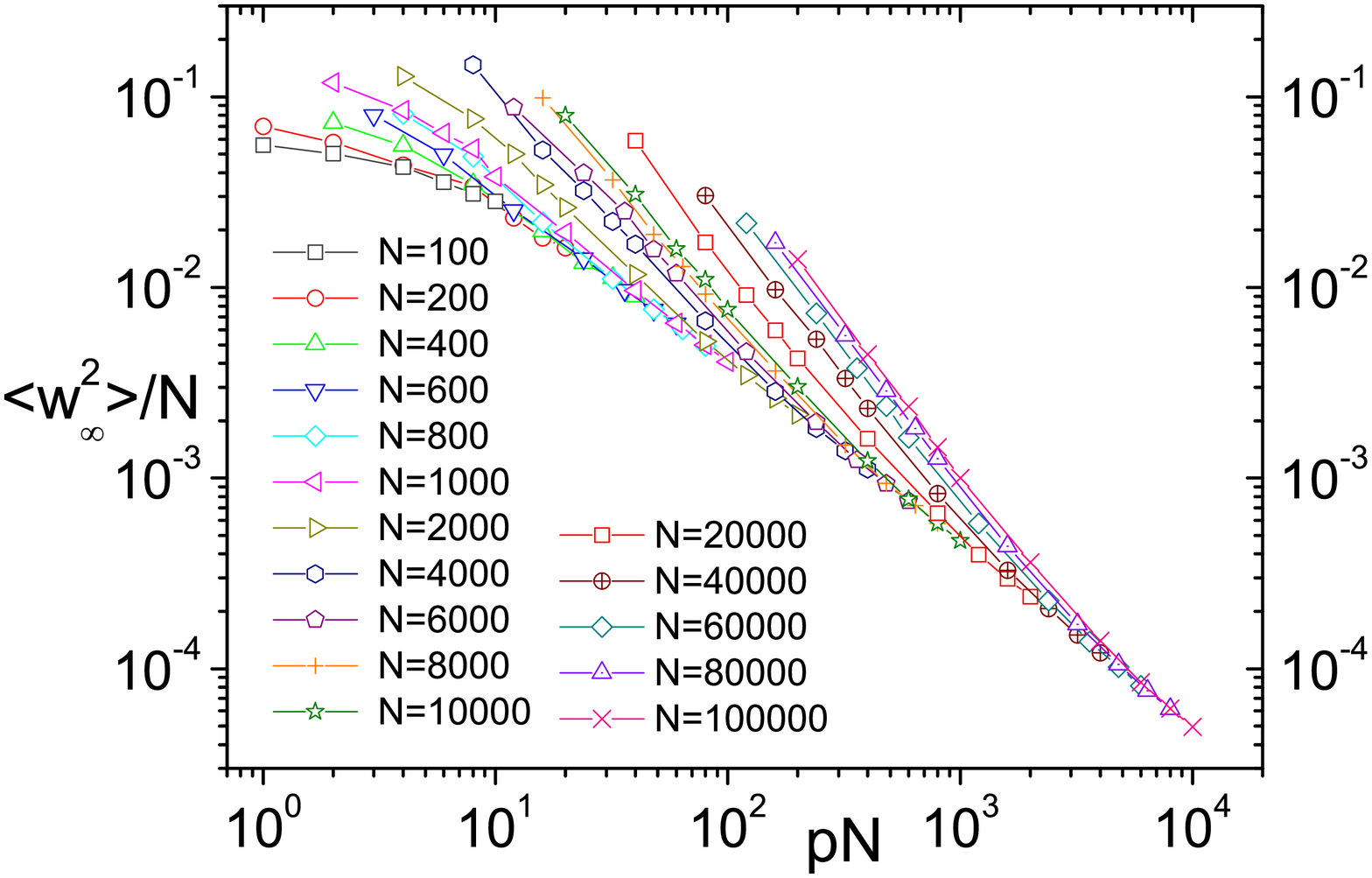}} a) {\em A--k1} \\
\end{minipage}
\vfill
\begin{minipage}[h]{1\linewidth}
\center{\includegraphics[width=1\linewidth]{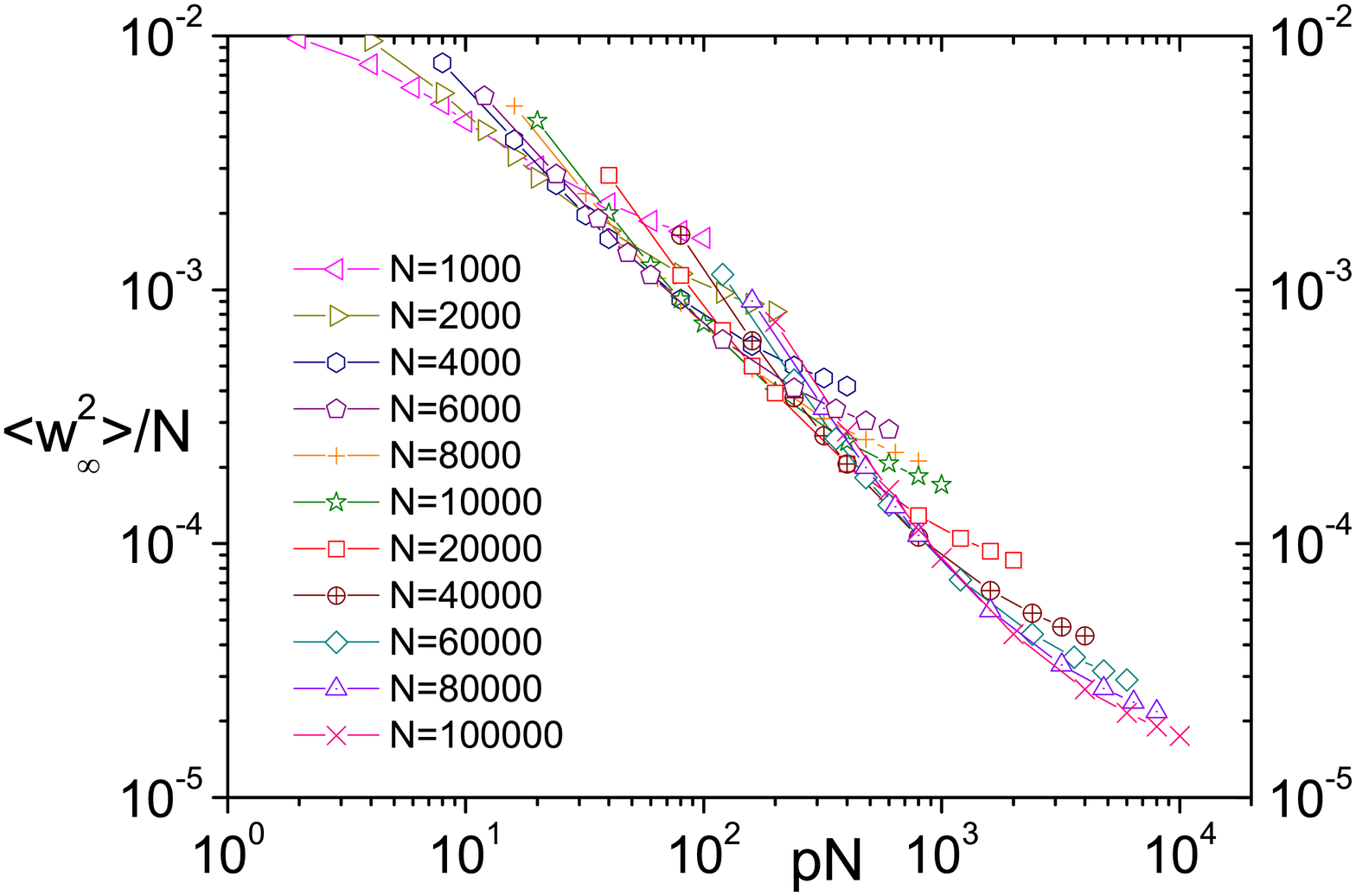}} \\b) {\em A--k2}
\end{minipage}
\vfill
\begin{minipage}[h]{1\linewidth}
\center{\includegraphics[width=1\linewidth]{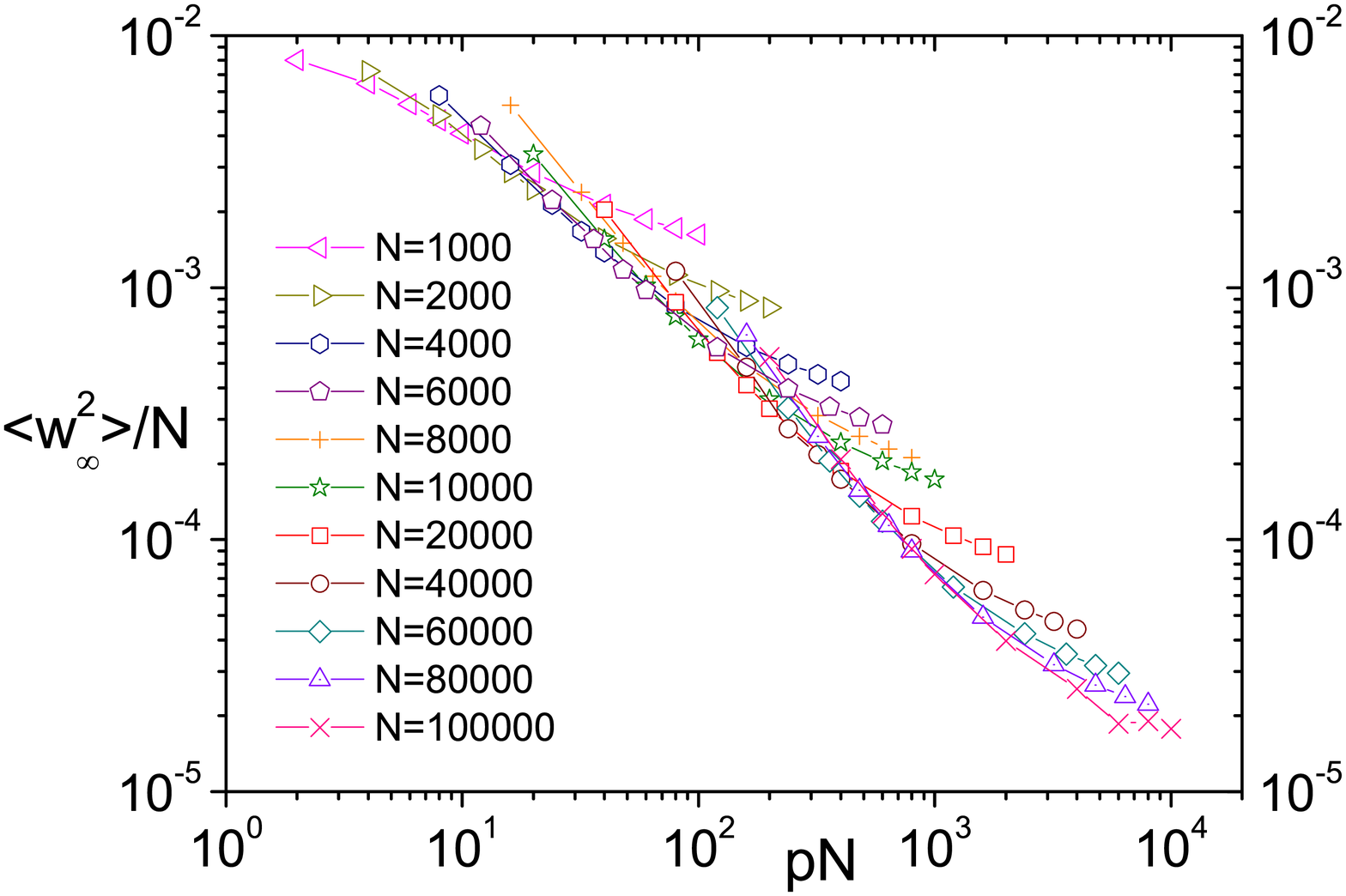}} c) {\em R--k2} \\
\end{minipage}
\vfill
\caption{(Color online) The steady-state width $\langle w^2_{\infty}\rangle$ normalized on
$N$ as a function of the number $pN$ of added or rewired links. The order of lines from top to bottom corresponds to the figure legend. }
\label{pic:w2normed}
\end{figure}

Figure~\ref{pic:w2normed} shows the collapse of the curves with a
normalized average width $\langle w^2_{\infty}\rangle/N$ as a function of
the normalized long-range links $pN$. The data collapse is good for the
networks {\em A--k2} and {\em R--k2} but rather poor for the network
{\em A--k1}. This is another argument that clustering affects properties
of the LVT evolution model for small $p$.

\section{Dependence on the local connectivity}
\label{sec:Local}

Real systems in the natural sciences often have large values of $k$. In this
section, we demonstrate how our results are sensitive to the variation of
$k$. We present a comparative study of the network properties and PDES
behavior for one-dimensional networks with the number of neighbors varying
from 2 to 16, i.e., for $k=1,2,4,8$. In our classification (see
Section~\ref{sec:SW}), they are respectively called {\em A--k1},
{\em A--k2}, {\em A--k4}, and {\em A--k8}.

\begin{figure}[ht!]
\center{\includegraphics[width=1\linewidth]{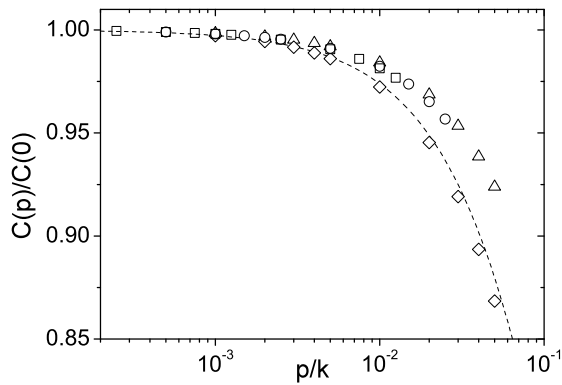}}
\caption{The normalized clustering coefficient of SW networks as a function of the
parameter $\tilde p=p/k$: triangles are {\em A--k2}, circles are
{\em A--k4}, squares are {\em A--k8}, diamonds are {\em R--k2}, and the
dashed line is equation~(\ref{eq:Newman}). Error bars are of the symbol
size.}
\label{pic:c(p)_dif_k}
\end{figure}

The variation of the clustering coefficient with the concentration $p$ of
long-range links is shown in Figure~\ref{pic:c(p)_dif_k}. It can be seen
that values of the normalized clustering coefficient coincide well for the
networks {\em A--k2}, {\em A--k4}, and {\em A--k8} (we recall that the value
of the clustering coefficient $C(0)$ for the network {\em A--k1} is zero).
For comparison, we plot the variation of the clustering coefficient for the
network {\em R--k2} and the corresponding approximation~(\ref{eq:Newman}).

The average shortest path $l$ is shown in Figure~\ref{l(p)_all_k} as a
function of the concentration of long-range links and in
Figure~{\ref{pic:l(pN)_diff_k} as a function of the system size. In all
cases, the behavior of $l$ for the presented range of $p$ and $N$ is well
approximated by Equation~(\ref{log-l-fit}). It is interesting that the
resulting fit in the values of $A$ varies slightly around the value
$A=0.30(1)$ and the values of $D$ are practically the same, $D=7.7(1)$, for
all investigated networks except {\em A-k1}, for which $D$ is much smaller,
$D=6.1(2)$.

\begin{figure}[ht!]
\center{\includegraphics[width=1\linewidth]{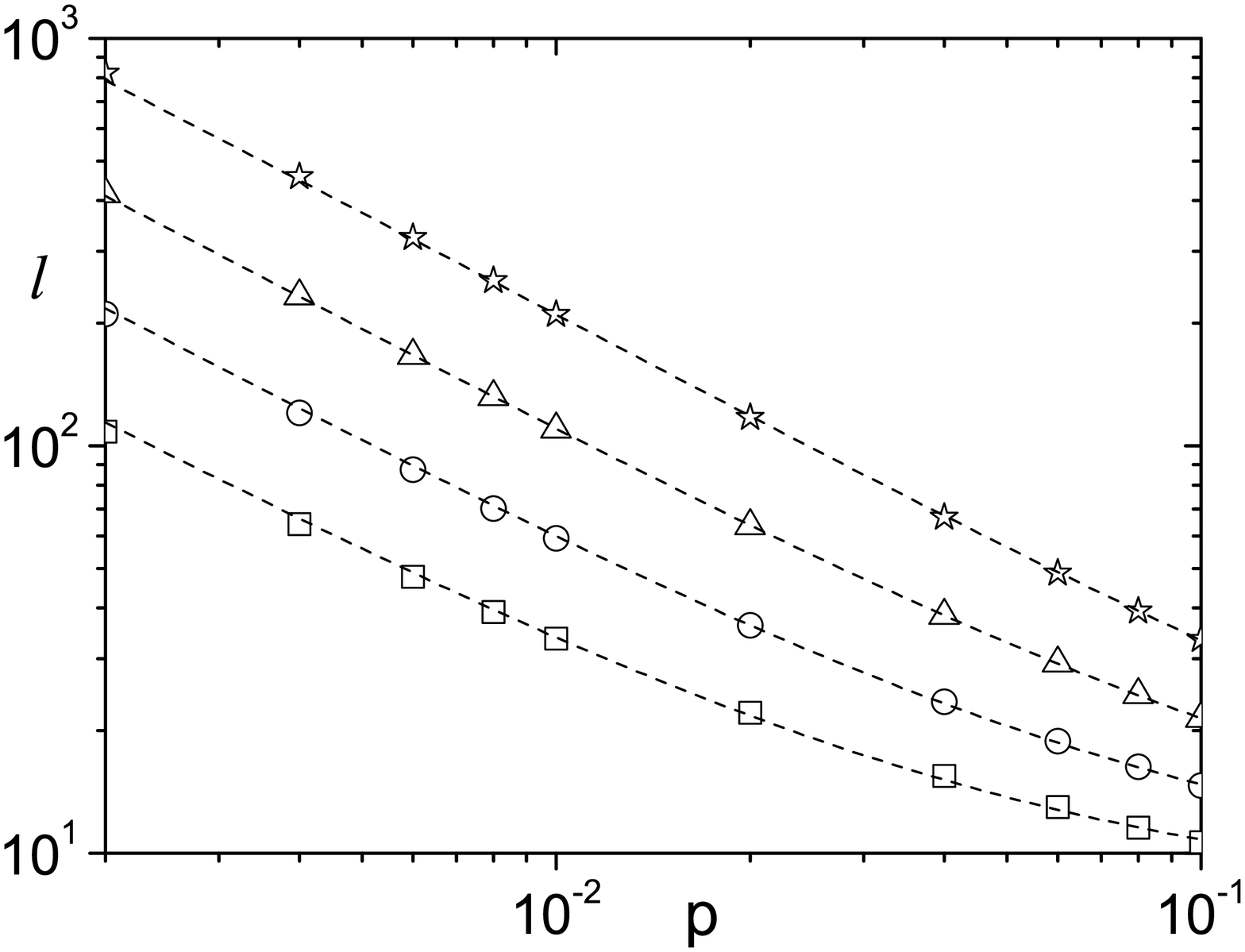}}
\caption{The average shortest path $l$ as a function of the parameter $p$
for systems of size $N=10^5$: stars are {\em A--k1}, triangles are
{\em A--k2}, circles are {\em A--k4}, squares are {\em A--k8}, and dashed
lines indicate fits using Equation~(\ref{log-l-fit}). Error bars are of the
symbol size.}
\label{l(p)_all_k}
\end{figure}

\begin{figure}[ht!]
\center
\center{\includegraphics[width=1\linewidth]{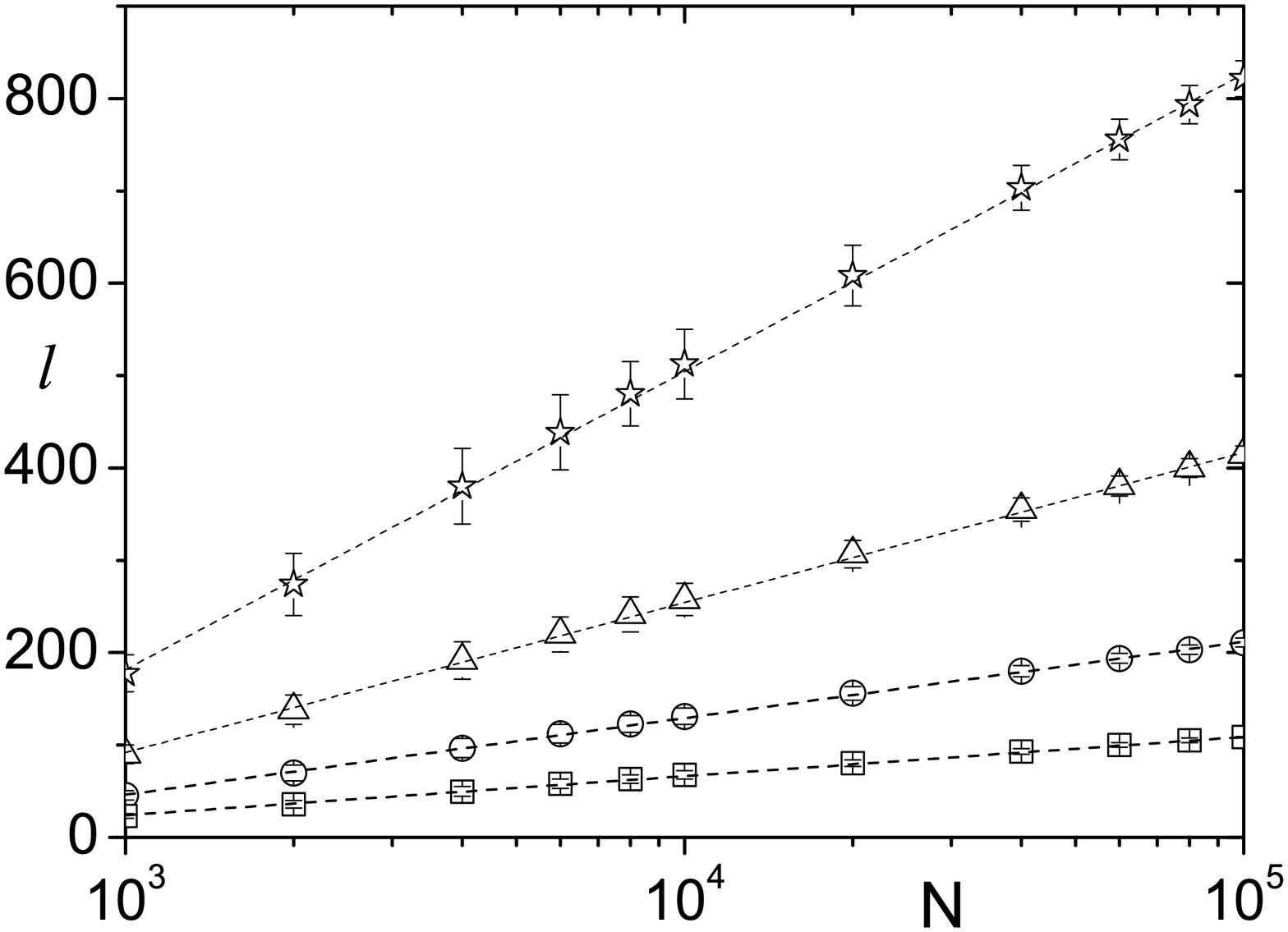}}\\
\caption{The average shortest path $l$ as a function of the number of nodes
for SW networks for $p=0.002$: stars are {\em A--k1},  triangles are
{\em A--k2}, circles are {\em A--k4}, squares are {\em A--k8}, and dashed
lines indicate fits using Equation~(\ref{log-l-fit}).}
\label{pic:l(pN)_diff_k}
\end{figure}

Simulation of the PDES on the SW networks with different $k$ leads to some
interesting observations. First, we found that the average speed
$\langle u\rangle$ of the time profile can be collapsed on one curve as a
function of the concentration $p$. Figures~\ref{pic:u_div_u0_(p_div_k)}
clearly show a good data collapse for $\langle u\rangle$ for the networks
{\em A--k2}, {\em A--k4}, and {\em A--k8} and rather poor collapse for the
network {\em A--k1}. Hence, a zero value of the clustering coefficient
$C(0)$ selects the network {\em A--k1} as a special case, while networks
with a nonzero value of $C(0)$ demonstrate a universal behavior. Values of
$\langle u_0\rangle$, i.e., the average speed on the time profile for the
network with $p=0$, are presented in Table~\ref{tab:u(k)}. Another
interesting observation is that $\langle u_0\rangle$ scales with $k$ as
$\langle u_0\rangle\propto k^{-0.84(1)}$. The data collapse shown in
Figure~\ref{pic:u_div_u0_(p_div_k)} can therefore be treated in the rescaled
variables $\left(\langle u\rangle\,k^{0.84}\right)$ and $(p/k)$.

The data collapse is even more nicely visible for the function $\Delta u$
normalized by dividing by $u_0$ (or, equivalently, multiplied by $k^{0.84}$)
as shown in Figure~\ref{pic:delta_u_div_u0(p_div_k)}. Therefore, the
exponent $B$ given by expressions~(\ref{eq:deltaU}) and (\ref{eq:B(N)}),
which characterize the behavior of $\Delta u\propto p^B$ for small values of
$p$ is indeed seems universal for $k=2,4$, and 8 with $B\approx 0.44(1)$.

\begin{figure}[ht!]
\center{\includegraphics[width=1\linewidth]{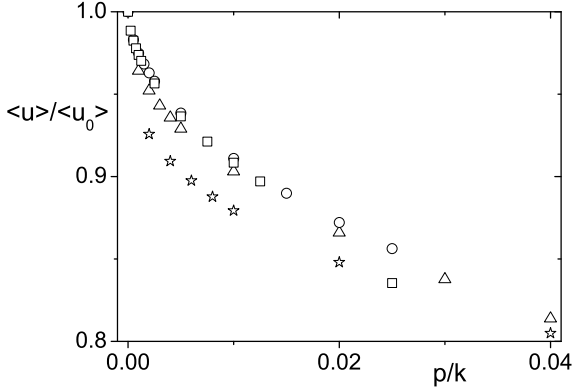}}
\caption{The average speed $\langle u\rangle$ divided by
$\langle u_{0}\rangle $ as a function of the parameter $\tilde p=p/k$ for
$N=10^5$ for network realizations: stars are {\em A--k1}, triangles are
{\em A--k2},  circles are {\em A--k4}, and squares are {\em A--k8}. Error
bars are of the symbol size.}
\label{pic:u_div_u0_(p_div_k)}
\end{figure}

\begin{figure}[ht!]
\center{\includegraphics[width=1\linewidth]{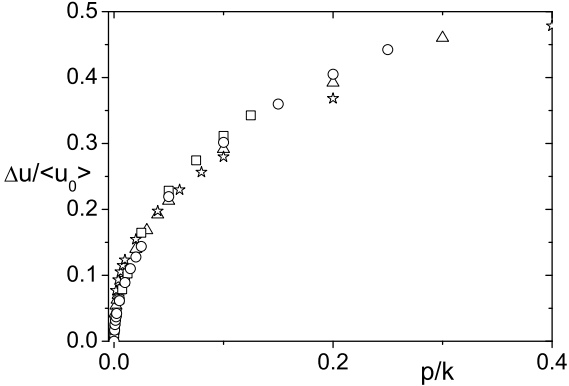}}
\caption{The difference between the speed $\Delta u=\langle u\rangle-
\langle u_{0}\rangle$ divided by $\langle u_{0}\rangle$ as a function of the
parameter $\tilde p=p/k$ for $N=10^5$ . Network realizations:  stars
are {\em A--k1}, triangles are {\em A--k2}, circles are {\em A--k4}, and
squares are {\em A--k8}. }
\label{pic:delta_u_div_u0(p_div_k)}
\end{figure}

\begin{table}[ht]
\begin{center}
\caption{Dependence of the average speed $\langle u_0\rangle$ in a regular
lattice ($p=0$) on the parameter $k$.}
\label{tab:u(k)}
\begin{tabular}{|l|l|}\hline
Network type&$\langle u_0\rangle$\\\hline
{\em A--k1}&0.246410(7)\\\hline
{\em A--k2}&0.14674(7)\\\hline
{\em A--k4}&0.08127(4)\\\hline
{\em A--k8}&0.04299(3)\\\hline
\end{tabular}
\end{center}
\end{table}

The behavior of the average width on regular ($p=0$) lattices {\em A--k1},
{\em A--k2}, {\em A--k4}, and {\em A--k8} demonstrates the same behavior
with the growth exponent of KPZ universality class $2\beta=2/3$ as shown in
Figure~\ref{pic:w2(t)_p0_diff_k}. The larger the value of $k$ is, the longer
the time required for entering the scaling regime.

Estimates of the values of $\beta$ as a function of $p$ are presented in
Table~\ref{tab:beta(p)_k4k8}. It can be seen that attaining the scaling
regime $t^{2/3}$ for $p=0$ is shifted to the larger network sizes.

In another words, translating our findings into computer science terms,
larger values of $k$ work in different directions: they suppress
desynchronization (width behavior), which is a positive sign, and suppress
utilization of processing time (average speed behavior).

\begin{figure}[ht!]
\center{\includegraphics[width=1\linewidth]{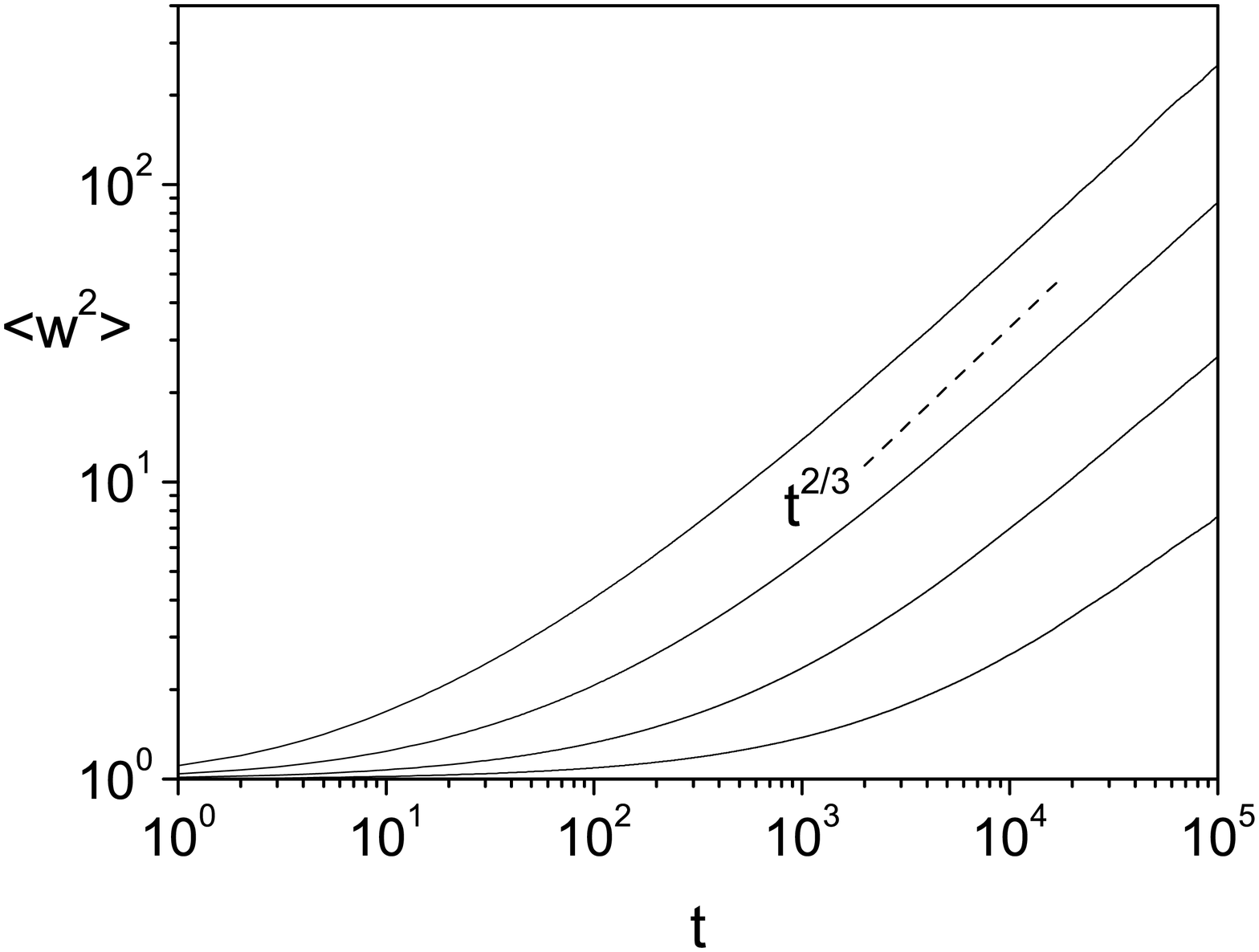}}
\caption{The average width $\langle w^2\rangle$ as a function of time on a
regular lattice with $p=0$ and different values of the parameter $k$ from
top to bottom: $k=1$, $k=2$, $k=4$, and $k=8$. The system size $N=10^5$. The
dashed line is a guide line for $t^{2/3}$.}
\label{pic:w2(t)_p0_diff_k}
\end{figure}

\begin{table}[ht]
\begin{center}
\caption{Dependence of the exponent $\beta$ on the concentration $p$ of
long-range links. }
\label{tab:beta(p)_k4k8}
\begin{tabular}{|l||l|l|l|l|}\hline
&\multicolumn{2}{c|}{\em A--k4}&\multicolumn{2}{c|}{\em A--k8}\\\hline
p&$N=10^4$&$N=10^5$&$N=10^4$&$N=10^5$\\\hline\hline
0&0.22(2)&0.291(1)&0.118(5)&0.25(1)\\\hline\hline
0.002&0.218(8)&0.27(1)&0.045(2)&0.06(1)\\\hline
0.004&0.148(4)&0.18(1)&0.0315(4)&0.040(2)\\\hline
0.006&0.113(2)&0.140(6)&0.027(2)&0.031(4)\\\hline
0.008&0.094(2)&0.114(5)&0.024(2)&0.027(1)\\\hline
0.01&0.083(3)&0.095(3)&0.022(2)&0.023(1)\\\hline
0.02&0.055(2)&0.059(4)&0.0158(3)&0.165(1)\\\hline
0.04&0.038(2)&0.039(2)&0.0113(3)&0.0116(4)\\\hline
0.06&0.032(1)&0.0313(5)&0.0104(5)&0.0098(5)\\\hline
0.08&0.0261(2)&0.0272(3)&0.0078(7)&0.0086(4)\\\hline
0.1&0.0240(2)&0.0239(3)&0.0069(4)&0.0074(1)\\\hline
\end{tabular}
\end{center}
\end{table}

\section{Conclusion}
\label{sec:Results}

We have investigated the influence of the SW communication topology on the
synchronization properties in the conservative PDES algorithm using a model
of the evolution of the LVT profile. We simulated the model on several SW
network realizations, which differed in their local properties and the
procedure for inserting the long-range links. The time evolution of the
model on a regular network (with only short-range interaction between PEs)
belongs to the KPZ universality class with the critical exponent values
$\alpha=1/2$ and $\beta=1/3$. In contrast, even a small number of long-range
links changes the behavior drastically. The growth exponent $\beta$ depends
logarithmically on the concentration of long-range links, $\beta\sim
-\ln{p}$, and the roughness exponent $\alpha$ drops to zero. The average
profile speed decreases as a power of the concentration $p$,
$\langle u \rangle=\langle u_0\rangle- \mathrm{const}\; p^B$ with $B=0.306(4)$ for the network
{\em A--k1}. It seems take universal value $B\approx 0.44(1)$ for the networks {\em A--k2}, {\em A--k4}, and
 {\em A--k8}, and it is $B=0.450(2)$ for the network {\em R--k2}. We found a data collapse of
the profile width as a function of the concentration $p$ for the two
realizations of the topology with a nonzero clustering coefficient. The
absence of data collapse for the network {\em A--k1} can probably be
attributed to the zero clustering coefficient. In other words, the network
{\em A--k1} is not quite a conventional SW network: it lacks clustering.

A model of time evolution for the conservative PDES was investigated
in~\cite{guclu2006synchronization, guclu2004small} for an underlying network
of the mean-field type where any site is connected by a single link to a
randomly chosen site and each site hence has exactly three links and each
non-neighbor link is activated with probability $p$. The results
in~\cite{guclu2006synchronization, guclu2004small} seem similar to some of
our results (we use the conventional SW network topology) but not all
results coincide. The common feature of the two approaches is that the
average shortest path grows logarithmically with the number of PEs. For
small $p$, it was found in~\cite{guclu2006synchronization} that the average
speed of the profile growth is
\begin{equation}
\langle u\rangle\simeq\frac{1}{4}+\frac{\sqrt{p}}{4\pi}-\mathcal{O}(p).
\end{equation}
In contrast, the average speed $\langle u\rangle$ in our simulations
decreases for any values of $p$ with a power-law dependence on $p$,
$\langle u\rangle\simeq\langle u_0\rangle-\mathrm{const}\,p^B$. The value
of the exponent $B$ is universal for the networks with a nonzero clustering
coefficient and takes a different value for the network with a zero
clustering coefficient.

The same qualitative behavior for the average profile speed
$\langle u\rangle$ and average profile width $\langle w^2_{\infty}\rangle$
was reported in~\cite{guclu2006synchronization, guclu2004small}, but the
dependence of the exponents $\alpha$ and $\beta$ on $p$ and of the average
speed $\langle u\rangle$ on the concentration $p$ of long-range links were
not analyzed.

We found that value of the clustering coefficient influences the progress
of the profile, and we argue that the larger the average coordination
number, the slower the profile speed.

In the language of computation processes, the results are as follows. First,
additional random long-range communication links in the communication
topology of PEs links cause more dependency checks during simulations and
reduce the average utilization of PEs, but the utilization remains positive,
i.e., the conservative PDES algorithm on SW networks remains free from
deadlock. Second, desynchronization becomes finite and decreases with the
amount of long-range communications. This enhances data collection and state
savings in PDES. The conservative synchronization algorithm of PDES hence
becomes \textit{fully scalable}: (1) the progress rate of simulations
remains positive, and (2) desyncronization of the LVT profile becomes finite
in the limit of a large number of PEs.

We compared the results on different SW network realizations. All have a
short average path, but they differ in the clustering property and the
construction method. One SW network has a zero clustering coefficient, and
the others are highly clustered. The highly clustered networks differ in
their construction (random long-range links were either added or rewired).
Qualitatively, the same results were obtained in all cases, i.e., the
communication network can be rearranged in any of the presented ways
to obtain a well-synchronized PDES algorithm. We found that the model
properties depend mainly on the number of long-range communication links and
weakly on the way the SW network is constructed. But there is no universal
behavior for all cases. For example, the average utilization decreases
faster with $p$ in the case of SW network with a zero clustering coefficient
compared with the SW networks with a high clustering coefficient.

A detailed analysis of the synchronization model of the conservative PDES
algorithm on the SW networks allows associating the parameters of the
considered model with the example of simulations of the particular models.
These in turn can shed light on how to optimize the simulations.

\acknowledgments

The work is done under the grant 14-21-00158 of the Russian Science
Foundation and in part (Section~\ref{sec:SW}) under the work plan
0236-2018-0001 of the Science Center in Chernogolovka. Detailed questions
from the anonymous referee led to including Section~\ref{sec:Local}.


\begin{thebibliography}{99}
\bibitem {foo1} The analogy with surface growth is not perfect in the case of SW topology, because we do not know any physical realizations of such long-range interactions.
\end{thebibliography}
\end{document}